\documentclass[
superscriptaddress,
preprint,
prd,tightenlines,showpacs,nofootinbib,
eqsecnum,
amsfonts,amsmath,amssymb]{revtex4-1}

\usepackage{bm}
\usepackage{graphicx} 
\usepackage{hyperref}
\usepackage{mathrsfs}
\usepackage{multirow}

\newcommand{\ud}{\mathrm{d}}
\newcommand{\uD}{\mathrm{D}}

\newcommand{\calO}{\mathcal{O}}

\newcommand{\ph}[1]{\phantom{#1}}

\usepackage{ulem}
\usepackage[usenames]{color}

\allowdisplaybreaks

\begin{document}

\title{Next-to-next-to-leading order spin-orbit effects in the
  \\gravitational wave flux and orbital phasing of compact binaries}

\author{Alejandro \textsc{Boh\'{e}}}\email{alejandro.bohe@uib.es}
\affiliation{Departament de F\'isica, Universitat de les Illes
  Balears, Crta.  Valldemossa km 7.5, E-07122 Palma, Spain}

\author{Sylvain \textsc{Marsat}}\email{marsat@iap.fr}
\affiliation{$\mathcal{G}\mathbb{R}\varepsilon{\mathbb{C}}\mathcal{O}$
  Institut d'Astrophysique de Paris --- UMR 7095 du CNRS,
  \ Universit\'e Pierre \& Marie Curie, 98\textsuperscript{bis}
  boulevard Arago, 75014 Paris, France}

\author{Luc \textsc{Blanchet}}\email{blanchet@iap.fr}
\affiliation{$\mathcal{G}\mathbb{R}\varepsilon{\mathbb{C}}\mathcal{O}$
  Institut d'Astrophysique de Paris --- UMR 7095 du CNRS,
  \ Universit\'e Pierre \& Marie Curie, 98\textsuperscript{bis}
  boulevard Arago, 75014 Paris, France}

\date{\today}

\begin{abstract}
We compute the next-to-next-to-leading order spin-orbit contributions
in the total energy flux emitted in gravitational waves by compact
binary systems. Such contributions correspond to the post-Newtonian
order 3.5PN for maximally spinning compact objects. Continuing our
recent work on the next-to-next-to-leading spin-orbit terms at 3.5PN
order in the equations of motion, we obtain the spin-orbit terms in
the multipole moments of the compact binary system up to the same
order within the multipolar post-Newtonian wave generation
formalism. Our calculation of the multipole moments is valid for
general orbits and in an arbitrary frame; the moments are then reduced to the
center-of-mass frame and the resulting energy flux is specialized to
quasi-circular orbits. The test-mass limit of our final result for the
flux agrees with the already known Kerr black hole perturbation
limit. Furthermore the various multipole moments of the compact binary
reduce in the one-body case to those of a single boosted Kerr black
hole. We briefly discuss the implications of our result for the
gravitational-wave flux in terms of the binary's phase evolution, and
address its importance for the future detection and parameter
estimation of signals in gravitational wave detectors.
\end{abstract}

\pacs{95.35.+d,95.36.+x,04.50.Kd}

\maketitle


\section{Introduction}
\label{Introduction}

Previous works \cite{MBFB12,BMFB13}\footnote{Hereafter these works
  will be referred to as Papers I \& II respectively.} have derived
the spin-orbit effects in the equations of motion of compact binary
systems (made of two black holes or neutron stars) at the
next-to-next-to-leading order beyond the dominant level. Such an order
corresponds to the post-Newtonian (PN) order 3.5PN $\sim 1/c^7$ in the
case of maximally spinning compact objects, \textit{i.e.} 2PN beyond
the leading spin-orbit effect at order 1.5PN $\sim 1/c^3$. The present
paper will continue Papers I \& II by investigating the gravitational
radiation field of compact binaries, notably the total gravitational
wave energy flux and orbital phase evolution, up to the same
next-to-next-to-leading 3.5PN level.

Including spin effects, and most importantly spin-orbit effects which
are linear in spins, in the templates of gravitational waves emitted
by compact binaries is of crucial importance for the accurate data
analysis of the advanced ground-based as well as future space-based
gravitational wave detectors. Astrophysical stellar-size black holes
\cite{AK01, Stroh01, Mcclint06, Gou11, Nowak12} as well as
super-massive black holes \cite{FM05, BrennR06, Brenn11} have spins,
and the spins will affect the gravitational waves of black hole
binaries through a modulation of their amplitude, phase and frequency
(including the precession of the orbital plane in the case of
non-aligned spins, see \textit{e.g.} Refs.~\cite{3mn, CF94, ACST94}).

The leading spin-orbit and spin-spin contributions in the equations of
motion have been obtained using various methods \cite{BOC75, BOC79,
  KWW93, K95, GR06, Porto06}; the next-to-leading corrections are also
known both for the spin-orbit \cite{TOO01, BBF06, DJSspin, Levi10,
  Porto10} and spin-spin terms \cite{HSS10, SHS08b, HS11ss, PR08a, PR08b};
next-to-next-to-leading spin-orbit corrections have been derived in
Refs.~\cite{HS11so, HSS13} and in Papers I \& II. Concerning the
radiation field of compact binaries, the leading spin-orbit and
spin-spin terms are known \cite{KWW93, K95}; the next-to-leading
spin-orbit terms at order 2.5PN $\sim 1/c^5$ were first obtained in
Ref.~\cite{BBF06} after a previous attempt in \cite{OTO98};
the 3PN $\sim 1/c^6$ spin-orbit contribution including the tail
integrals were computed in \cite{BBF11}, after intermediate results at
the same order (but including spin-spin terms) were given in
\cite{PRR10}; finally, the next-to-next-to-leading order contributions
in the multipole moments and the energy flux, corresponding to 3.5PN
$\sim 1/c^7$, is the topic of the present paper and has never been
addressed before.

Following the previous investigations \cite{BBF06,BBF11} we
shall apply the so-called multipolar post-Newtonian approach to
gravitational radiation. This approach has been extensively developped
over the years (see \cite{Bliving} for a review). It combines a mixed
post-Minkowskian and multipolar expansion for the gravitational field
in the exterior of a general source having compact support~\cite{BD86,
  B87, BD92, B93}, with a matching to the post-Newtonian expansion of
the inner field in the near zone of a post-Newtonian
source~\cite{BD89, B95, B98tail, B98mult, PB02}. The gravitational
waveform and various fluxes like the energy flux (or gravitational
``luminosity'') are expanded in a series of radiative multipole
moments, that are then related to appropriate source-rooted multipole
moments, expressed as integrals over the matter and gravitational
fields in the source. The time derivatives of the multipole moments
are performed using the equations of motion of the source, which must
thus be known beforehand with the same accuracy as the one aimed for in
the radiation field.

Though the formalism can be applied to any post-Newtonian source, it
does require a model for the source. In the case of compact binaries,
the compact objects are described by point-like particles
characterized only by their masses and their spins. The appropriate
model in this context is an effective ``pole-dipole'' description,
based on a stress-energy tensor made of a monopole or mass term
involving delta functions, and a dipole or spin term made of gradients
of delta functions. The pole-dipole description of spinning particles
has been developed by many authors \cite{Math37, Papa51, Papa51spin,
  CPapa51spin, Tulc1, Tulc2, Traut58, Dixon64, Dixon73, Dixon79,
  BI80}. The model is to be suplemented by an ultra-violet (UV)
regularization in order to remove the infinite self-field of the point
masses, see \textit{e.g.} \cite{BFreg,BDE04}. In Papers I \& II and
the present paper, we use (and present justifications for that use at
the aimed level of accuracy) the Hadamard ``partie finie''
regularization \cite{Hadamard}, together with the Gel'Fand-Shilov
prescription for distributional derivatives \cite{gelfand}, equivalent
to Schwartz distributional derivatives \cite{Schwartz}. Explicit
expressions for the waveform and flux can then be obtained in terms of
the source's positions and velocities, and these can finally be
converted into gauge-invariant quantities that are directly in use for
building the gravitational wave templates.

The plan of this paper is as follows. The Section~\ref{II} is devoted
to the general wave generation formalism and the Section~\ref{III} to
the application to spinning compact binaries. We give the required
formulas in Sec.~\ref{IIA} for the various types of multipole moments
(radiative, canonical and source), and in Sec.~\ref{IIB} for the
general post-Newtonian solution we employ. The complete results for the
source multipole moments at next-to-next-to-leading spin-orbit level
are presented in Sec.~\ref{IIIA}, while we obtain in Sec.~\ref{IIIB}
the next-to-next-to-leading energy flux and orbital phase
evolution. We also present a numerical estimate of the new terms in Sec. \ref{IIIB}. In Appendix~\ref{appA} we give alternative expressions of
the source terms to be inserted into the source multipole moments. In
Appendix~\ref{appB} we check the agreement with the so-called boosted
Kerr black hole limit, obtained when the mass and spin of one of the
two black holes are set to be exactly zero.

\section{Gravitational wave generation formalism}
\label{II}

\subsection{Radiative and source multipole moments}
\label{IIA}

The gravitational waveform, generated by an isolated source described
by a stress-energy tensor with compact support, is the
transverse-tracefree (TT) projection of the metric deviation, say
$h_{ij}^\mathrm{TT}\equiv(g_{ij}-\delta_{ij})^\mathrm{TT}$. It is
defined in a suitable radiative coordinate system
$X^\mu=(c\,T,\mathbf{X})$, at the leading-order $1/R$ when the distance
$R=\vert\mathbf{X}\vert$ to the source tends to infinity, with the
retarded time $T_R\equiv T-R/c$ being fixed. In radiative coordinates
the retarded time $T_R$ asymptotically coincides with a null
coordinate. The waveform reads\footnote{We denote by $L=i_1\cdots
  i_\ell$ a multi-index composed of $\ell$ multipolar spatial indices
  $i_1, \cdots, i_\ell$ ranging from 1 to 3. Similarly $L-1=i_1\cdots
  i_{\ell-1}$ and $kL-2=k i_1\cdots i_{\ell-2}$; $N_L = N_{i_1}\cdots
  N_{i_\ell}$ is the product of $\ell$ spatial vectors $N_i$. In the
  case of summed-up (dummy) multi-indices $L$, we do not write the
  $\ell$ summations from 1 to 3 over their indices. The
  transverse-traceless (TT) projection operator is denoted
  $\mathcal{P}^\mathrm{TT}_{ijkl} =
  \mathcal{P}_{ik}\mathcal{P}_{jl}-\frac{1}{2}\mathcal{P}_{ij}\mathcal{P}_{kl}$
  where $\mathcal{P}_{ij}=\delta_{ij}-N_iN_j$ is the projector
  orthogonal to the unit direction $\mathbf{N}=\mathbf{X}/R$ of the radiative
  coordinate system $X^\mu=(c\,T,\mathbf{X})$. The quantity
  $\varepsilon_{ijk}$ is the Levi-Civita antisymmetric symbol such
  that $\varepsilon_{123}=1$. The symmetric-trace-free (STF)
  projection is indicated using brackets or a hat. Thus
  $U_L=\hat{U}_L=U_{\langle L\rangle}$ and $V_L=\hat{V}_L=V_{\langle
    L\rangle}$ for STF moments. We denote time derivatives with a
  superscript $(n)$, and we indicate the symmetrization operation with
  round parentheses.}
\begin{align}\label{waveform}
h^\mathrm{TT}_{ij} = \frac{4G}{c^2R}
\,\mathcal{P}^\mathrm{TT}_{ijkl}(\mathbf{N})
\sum^{+\infty}_{\ell=2}\frac{N_{L-2}}{c^\ell\ell !} \biggl[
  U_{klL-2}(T_R) - \frac{2\ell}{c(\ell+1)}\,N_{m} \,\varepsilon_{mn(k}
  \,V_{l)nL-2}(T_R) \biggr] + \mathcal{O}\left(\frac{1}{R^2}\right)\,.
\end{align}
The waveform is parametrized by two sets of symmetric and trace-free
(STF) multipole moments, $U_L$ of mass type and $V_L$ of current type,
which constitute the observables of the gravitational wave at infinity
and are called the radiative moments \cite{Th80}. They are functions
of the retarded time $T_R$ in the radiative coordinate
system. Plugging Eq.~\eqref{waveform} into the standard expression for
the gravitational-wave energy flux we get~\cite{Th80}
\begin{equation}\label{flux}
\mathcal{F} = \sum_{\ell = 2}^{+ \infty} \frac{G}{c^{2\ell
    +1}}\,\biggl[ \frac{(\ell+1)(\ell+2)}{(\ell-1) \ell \, \ell!
    (2\ell+1)!!} U_L^{(1)} U_L^{(1)} + \frac{4\ell (\ell+2)}{c^2
    (\ell-1) (\ell+1)!  (2\ell+1)!!} V_L^{(1)} V_L^{(1)}\biggr]\,.
\end{equation}

The radiative moments $U_L$ and $V_L$ are then related to some
specific source-rooted multipole moments as follows. To implement the
non-linearities in the propagation of the gravitational waves from the
source to infinity, we express them as some non-linear functionals,
which can in principle be developed at any order, of some
``canonical'' moments $M_L$ and $S_L$. When developped at the 1.5PN
order they display the effect of tails and read~\cite{BD92,B95}
\begin{subequations} \label{tails}
\begin{align}
U_L(T_R) &= M_L^{(\ell)} + \frac{2 G M}{c^3} \int_{-\infty}^{T_R}\! \ud t \,
M_L^{(\ell +2)}(t) \biggl[\ln \biggl(\frac{T_R-t}{2\tau_0} \biggr)+
  \kappa_\ell \biggr] 
+\mathcal{O}\Bigl(\frac{1}{c^5}\Bigr)\, ,\label{eq:tailsU}\\ V_L(T_R) &=
S_L^{(\ell)} + \frac{2 G M}{c^3} \int_{-\infty}^{T_R} \!  \ud t \,
S_L^{(\ell +2)}(t) \biggl[\ln \biggl(\frac{T_R-t}{2\tau_0} \biggr)+
  \pi_\ell \biggr]
+\mathcal{O}\Bigl(\frac{1}{c^5}\Bigr)\, ,\label{eq:tailsV}
\end{align}
\end{subequations}
where $M$ is the mass monopole or Arnowitt-Deser-Misner (ADM) total
mass. The quantities $\kappa_\ell$ and $\pi_\ell$ denote some
numerical rational fractions and $\tau_0$ is an arbitrary constant
time scale; we shall not need any of these here.

Next the canonical moments $M_L$ and $S_L$ themselves are given as
some non-linear functionals of the ``source'' moments $I_L$ and $J_L$,
and also of four supplementary ``gauge'' moments $W_L$, $X_L$, $Y_L$
and $Z_L$. In general, the canonical and source moments agree up to
the 2.5PN order, namely
\begin{subequations}\label{MLSL}\begin{align}
M_L &= I_L + \mathcal{O}\left(\frac{1}{c^5}\right)\,,\\ 
S_L &= J_L + \mathcal{O}\left(\frac{1}{c^5}\right)\,.
\end{align}\end{subequations}
Since we address the computation of the spin-orbit 3.5PN contribution
to the energy flux \eqref{flux}, we see that we only have to consider
a possible spin-orbit $1/c^7$ term in the mass quadrupole (since spin contributions add at least a factor $1/c$). The
relation between $M_{ij}$ and $I_{ij}$ is given by (see \textit{e.g.}
\cite{BFIS08}):
\begin{equation}\label{MijIij}
	M_{ij} = I_{ij} + \frac{4G}{c^{5}} \left[ W^{(2)}I_{ij} -
          W^{(1)}I_{ij}^{(1)}\right] +
        \calO\left(\frac{1}{c^{7}}\right) \,.
\end{equation}
As was already noticed in Ref.~\cite{BBF11}, the leading order spin
contributions to $W$ and $I_{ij}$ both start at $\calO(c^{-3})$, so
that $(M_{ij})_{S}=(I_{ij})_{S}+\calO(c^{-8})$. For our purposes, we
can therefore ignore the distinction between canonical $M_L$, $S_L$
and source $I_L$, $J_L$ moments.

Finally the source multipole moments are defined for a general
post-Newtonian matter source for any multipolar order $\ell\geq 2$,
and up to any post-Newtonian order. They are explicitly given by
\cite{B98mult}\footnote{The brackets surrounding indices denote the
  STF projection; the STF product of $\ell$ spatial vectors is written
  as $\hat{x}_L \equiv x_{\langle i_1}\cdots x_{i_\ell\rangle} \equiv
  \text{STF}[x_L]$.}
\begin{subequations}\label{ILJL}\begin{eqnarray}
I_L(t)&=& \mathop{\mathrm{FP}}_{B=0}\,\int
\ud^3\mathbf{x}\,\left(r/r_0\right)^B \int^1_{-1} \ud z\left\{
\delta_\ell\,\hat{x}_L\,\Sigma
-\frac{4(2\ell+1)}{c^2(\ell+1)(2\ell+3)} \,\delta_{\ell+1}
\,\hat{x}_{iL} \,\Sigma_i^{(1)}\right.\nonumber\\ &&\qquad\quad
\left. +\frac{2(2\ell+1)}{c^4(\ell+1)(\ell+2)(2\ell+5)}
\,\delta_{\ell+2}\,\hat{x}_{ijL}\Sigma_{ij}^{(2)}\right\}
(\mathbf{x},t+z\,r/c)\,,\label{IL}\\
J_L(t)&=& \mathop{\mathrm{FP}}_{B=0}\,\varepsilon_{ab<i_\ell} \int
\ud^3 \mathbf{x}\,\left(r/r_0\right)^B \int^1_{-1} \ud z\biggl\{
\delta_\ell\,\hat{x}_{L-1>a} \,\Sigma_b \nonumber\\ &&\qquad\quad
-\frac{2\ell+1}{c^2(\ell+2)(2\ell+3)}
\,\delta_{\ell+1}\,\hat{x}_{L-1>ac} \,\Sigma_{bc}^{(1)}\biggr\}
(\mathbf{x},t+z\,r/c)\,,\label{JL}
\end{eqnarray}\end{subequations}
The finite part operation FP in front represents an infra-red (IR)
regularization defined by analytic continuation in a complex parameter
$B$, and involves the same arbitrary scale $r_0=c\,\tau_0$ as in
Eqs.~\eqref{tails}, which will be irrelevant for the present work.\footnote{This scale enters the relation between the retarded time in radiative coordinates and the one in source-rooted harmonic coordinates: $T_R=t-\frac{r}{c}-\frac{2GM}{c^3}\ln\left(\frac{r}{r_0}\right)$.}

The basic building ``blocks'' $\Sigma$, $\Sigma_i$ and $\Sigma_{ij}$
entering the latter formulas are evaluated at the position $\mathbf{x}$
and at time $t+z\,r/c$ in a harmonic coordinate system $(t,\mathbf{x})$
covering the source (where $r=\vert\mathbf{x}\vert$). They are defined
by
\begin{equation}\label{Sigma}
\Sigma \equiv \frac{\tau^{00} +\tau^{ii}}{c^2}\,,\qquad\Sigma_i \equiv
\frac{\tau^{0i}}{c}\,,\qquad\Sigma_{ij} \equiv\tau^{ij}\,,
\end{equation} 
together with $\tau^{ii}\equiv \delta_{ij}\tau^{ij}$. Here
$\tau^{\mu\nu}$ denotes the \textit{post-Newtonian expansion} of the
total pseudo stress-energy tensor of the matter and gravitational
fields, say
\begin{equation}\label{tau}
\tau^{\mu\nu} \equiv \text{PN}\Bigl[\,\vert g\vert
  T^{\mu\nu}+\frac{c^4}{16\pi G}\Lambda^{\mu\nu}(h)\Bigr]\,,
\end{equation}
where $T^{\mu\nu}$ is the stress-energy tensor of the matter source,
and $\Lambda^{\mu\nu}(h)$ represents the gravitational source term
which is given by a complicated non-linear, quadratic at least,
functional of the field variable $h^{\mu\nu}$ and its first and second
space-time derivatives. The pseudo-tensor appears in the
right-hand-side of the Einstein field equations, when ``relaxed'' by
the condition of harmonic (or de Donder) coordinates.\footnote{The
  post-Newtonian expansion of the relaxed Einstein field equations
  takes the form $\Box h^{\mu\nu}=\frac{16\pi
    G}{c^4}\tau^{\mu\nu}$, where
  $\Box\equiv\eta^{\rho\sigma}\partial_{\rho\sigma}$ is the flat
  space-time d'Alembertian operator. Here
  $h^{\mu\nu}\equiv\sqrt{-g}\,g^{\mu\nu}-\eta^{\mu\nu}$, where
  $g^{\mu\nu}$ is the inverse and $g$ the determinant of the usual
  covariant metric $g_{\mu\nu}$; $\eta^{\mu\nu}$ is an auxiliary
  Minkowskian metric, $\eta^{\mu\nu}=\mathrm{diag}(-1,1,1,1)$. The
  harmonic-coordinate condition reads $\partial_\nu
  h^{\mu\nu}=0$. Note that the conservation of the pseudo tensor,
  $\partial_\nu\tau^{\mu\nu} = 0$, is the consequence of the
  harmonic-coordinate condition.} The expressions \eqref{ILJL} involve
an intermediate integration over the variable $z$, with associated
weighting function
\begin{subequations}\label{deltal}\begin{eqnarray}
\delta_\ell (z) &\equiv& \frac{(2\ell+1)!!}{2^{\ell+1} \ell!}
\,(1-z^2)^\ell\,,\qquad\int_{-1}^{1}\ud z\,\delta_\ell (z) = 1\,.
\end{eqnarray}\end{subequations}
In practice the post-Newtonian expansion of the source moments
\eqref{ILJL} is performed by means of the formal infinite series
\begin{equation}\label{intdeltal}
\int^1_{-1} dz~ \delta_\ell(z) \,\Sigma(\mathbf{x},t+z\,r/c) =
\sum_{k=0}^{+\infty}\,\frac{(2\ell+1)!!}{(2k)!!(2\ell+2k+1)!!}
\,\left(\frac{r}{c}\right)^{2k}\!\Sigma^{(2k)}(\mathbf{x},t)\,.
\end{equation}

\subsection{Explicit solution for the post-Newtonian metric}
\label{IIB}

To get explicit results at a given post-Newtonian order we need a
solution of the relaxed Einstein field equations. As in Paper I, we
parametrize an explicit solution by means of a set of retarded
potentials denoted by $V$, $V_i$, $\hat{W}_{ij}$, $\hat{R}_{i}$,
$\hat{X}$, $\hat{Z}_{ij}$. Here, contrarily to our previous work on
the equations of motion, we will not need the highest-order potentials
$\hat{Y}_i$ and $\hat{T}$ that enter respectively $g_{0i}$ at
$\calO(c^{-7})$ and $g_{00}$ at $\calO(c^{-8})$. All these potentials
are ``Newtonian'' in the sense that they admit a finite non-zero limit
when $c\to +\infty$. They enter the components of the usual covariant
metric as follows:
\begin{subequations}\label{metricg}
\begin{align} 
g_{00} &= -1 + \frac{2}{c^{2}}V - \frac{2}{c^{4}} V^{2} +
\frac{8}{c^{6}} \left(\hat{X} + V_{i} V_{i} +
\frac{V^{3}}{6}\right)+\calO\left(\frac{1}{c^{8}}\right)\;,\\ 
g_{0i} & = - \frac{4}{c^{3}} V_{i} - \frac{8}{c^{5}}
\hat{R}_{i} +
\calO\left(\frac{1}{c^{7}}\right)\;,\\ 
g_{ij} & = \delta_{ij} \left[1 + \frac{2}{c^{2}}V + \frac{2}{c^{4}}
  V^{2} + \frac{8}{c^{6}} \left(\hat{X} + V_{k} V_{k} +
  \frac{V^{3}}{6}\right)\right] \nonumber\\ & +
\frac{4}{c^{4}}\hat{W}_{ij} + \frac{16}{c^{6}} \left( \hat{Z}_{ij} +
\frac{1}{2} V \hat{W}_{ij} - V_{i} V_{j} \right) +
\calO\left(\frac{1}{c^{8}} \right) \;.
\end{align}
\end{subequations}
Equivalently, and more useful for the present work, they enter the ``gothic" metric as:
\begin{subequations}\label{metrich}
\begin{align} 
& \frac{h^{00}+h^{ii}}{2} = -\frac{2}{c^{2}}V - \frac{4}{c^{4}} V^{2} -
\frac{8}{c^{6}} \left(\hat{X} + \frac{1}{2} V \hat{W} +
\frac{2}{3}V^{3}\right) +\calO\left(\frac{1}{c^{8}} \right)\;,\\ 
& h^{0i} = - \frac{4}{c^{3}} V_{i} - \frac{8}{c^{5}} \left(\hat{R}_{i} + V
V_i\right)  +
\calO\left(\frac{1}{c^7}\right)\;,\\ 
& h^{ij} = - \frac{4}{c^{4}}\left(\hat{W}_{ij} - \frac{1}{2}
\delta_{ij} \hat{W}\right) - \frac{16}{c^{6}} \left( \hat{Z}_{ij} -
\frac{1}{2} \delta_{ij} \hat{Z} \right) +
\calO\left(\frac{1}{c^8}\right) \;.
\end{align}
\end{subequations}
Each of these potentials is a retarded solution of a flat space-time
wave equation sourced by matter densities components and appropriate
lower order potentials. The matter densities are defined from the
components of the matter stress-energy tensor by
\begin{equation}\label{sourcedensity}
\sigma \equiv \frac{T^{00} +T^{ii}}{c^2}\,,\qquad\sigma_i \equiv
\frac{T^{0i}}{c}\,,\qquad\sigma_{ij} \equiv T^{ij}\,.
\end{equation} 
Denoting with $\Box_{R}^{-1}S$ the usual retarded flat d'Alembertian
integral, \textit{i.e.} the retarded solution of
$\Box\phi\equiv\eta^{\mu\nu}\partial_{\mu\nu}\phi = S$, the
latter potentials are defined by
{\allowdisplaybreaks
\begin{subequations}\label{defpotentials}
\begin{align}
V &= \Box_{R}^{-1}[-4 \pi G\, \sigma]\,,\label{V}\\
V_{i}
&= \Box_{R}^{-1}[-4 \pi G\, \sigma_{i}]\,,\\ 
\hat{X} &=
\Box_{R}^{-1}\left[\vphantom{\frac{1}{2}} -4 \pi G\, V
  \sigma_{ii} + \hat{W}_{ij} \partial_{ij} V + 2 V_{i} \partial_{t}
  \partial_{i} V + V \partial_{t}^{2} V\right.\nonumber\\ &
  \qquad\quad\left. + \frac{3}{2}(\partial_{t} V)^{2} - 2 \partial_{i}
  V_{j}\partial_{j} V_{i}\right]\,,\\ 
\hat{R}_{i} & =
\Box_{R}^{-1}\left[-4 \pi G\, (V \sigma_{i} - V_{i} \sigma)
  - 2 \partial_{k} V \partial_{i} V_{k} - \frac{3}{2} \partial_{t} V
  \partial_{i} V\right]\,, \\ 
\hat{W}_{ij} & = \Box_{R}^{-1}\left[-4 \pi G\, (\sigma_{ij} -
  \delta_{ij} \sigma_{kk}) - \partial_{i} V \partial_{j}
  V\right]\,,\label{Wij}\\
\hat{Z}_{ij} & = \Box_{R}^{-1} \left[\vphantom{\frac{1}{2}}-4 \pi G\,
  V \left(\sigma_{ij} - \delta_{ij} \sigma_{kk}\right) - 2
  \partial_{(i} V \partial_{t} V_{j)} + \partial_{i} V_{k}
  \partial_{j} V_{k} + \partial_{k} V_{i}\partial_{k} V_{j}
  \right. \nonumber\\ &\qquad\quad \left. - 2 \partial_{(i} V_{k}
  \partial_{k} V_{j)} - \delta_{ij} \partial_{k} V_{m} (\partial_{k}
  V_{m} - \partial_{m} V_{k}) - \frac{3}{4} \delta_{ij} (\partial_{t}
  V)^{2} \right]\,.
\end{align}
\end{subequations}
}\noindent
With the latter explicit post-Newtonian solution in hands one obtains
the basic buiding blocks \eqref{Sigma} entering the source multipole
moments \eqref{ILJL} as
{\allowdisplaybreaks
\begin{subequations}
\begin{eqnarray}
\Sigma &=& \biggl[
1
+\frac{4V}{c^2}
+\frac{4}{c^4} (2V^2+\hat{W})
+\frac{16}{c^6}\left(V_iV_i+\frac{2}{3}V^3+V \hat{W}+\hat{X}+\hat{Z}\right)
\biggr]\sigma 
 - \frac{1}{\pi Gc^2}\,\partial_i V \partial_i V \nonumber\\
&&
+ \frac{1}{\pi Gc^4} \biggl\{ 
-\frac{1}{2} (\partial_t V)^2
- 2V_i\partial_t \partial_i V
-V \partial^2_t V
- \frac{7}{2} V\partial_i V\partial_i V
- \hat{W}_{ij}\partial^2_{ij} V
-\partial_i\hat{W}\partial_i V\nonumber\\
&& \qquad\quad+2 \partial_i V_j \partial_j V_i
+\frac{1}{2}\hat{W}\partial_{ii} V
\biggr\}\nonumber\\
&&+\frac{1}{\pi Gc^6}\biggl\{ 
-\partial_t V \partial_t \hat{W}
-\frac{7}{2}(\partial_t V)^2 V
+2\partial_t V_i \partial_t V_i
-4 \hat{R}_i \partial_t \partial_i V
-12 V V_i \partial_t \partial_i V\nonumber\\
&& \qquad\quad
- 6 V^2 \partial_t^2 V
-\frac{1}{2} \hat{W} \partial_t^2 V
- 6 V_i \partial_t V \partial_i V
+ 4 V \partial_t V_i \partial_i V
-7 V^2 \partial_i V \partial_i V\nonumber\\
&& \qquad\quad
-8 \partial_i \hat{X}\partial_i V
- \hat{W}_{ij} \partial_i V \partial_j V
-2 \partial_t \hat{W} \partial_i V_i
-4 V \hat{W}_{ij} \partial_{ij} V
- 4 \hat{Z}_{ij} \partial_{ij} V\nonumber\\
&& \qquad\quad
+ 8 \partial_i V_j \partial_j \hat{R}_i
-\frac{3}{2} \hat{W} \partial_i V \partial_i V
- 8 V_i \partial_j V_i \partial_j V
- 4 V \partial_i \hat{W} \partial_i V
- 4 \partial_i \hat{Z} \partial_i V\nonumber\\
&& \qquad\quad
+ 4 \partial_t \hat{W}_{ij} \partial_j V_i
+ 6 V \partial_i V_j \partial_j V_i
+ 2 V \partial_j V_i \partial_j V_i
+ 2 V \hat{W} \partial_{ii} V
+ 2 \hat{Z} \partial_{ii} V\nonumber\\
&& \qquad\quad
+ \partial_i \hat{W}_{jk} \partial_i \hat{W}_{jk}
-\frac{1}{2} \partial_i \hat{W} \partial_i \hat{W}
\biggr\} +\calO\left(\frac{1}{c^8}\right)\,, \\
\Sigma_i &=& \biggl[ 
1 +\frac{4V}{c^2}+\frac{4}{c^4} (2V^2+\hat{W})
\biggr] \sigma_i
+ \frac{1}{\pi Gc^2} \left\{ 
\partial_k V(\partial_i V_k -\partial_k V_i) +
\frac{3}{4} \partial_t V \partial_i V 
\right\} \nonumber \\ 
&&+\frac{1}{\pi Gc^4}\biggl\{
\partial_t V \partial_t V_i
- 2 V_j \partial_t \partial_j V_i
-V \partial_t^2 V_i
-2 \partial_j V \partial_j \hat{R}_i
+\partial_t \hat{W}_{ij} \partial_j V
-\frac{3}{2} V_i \partial_j V \partial_j V\nonumber\\
&& \qquad\quad
-2 V \partial_j V_i \partial_j V
-\hat{W}_{jk} \partial_{jk} V_i
+ \partial_j \hat{W}_{ik} \partial_k V_j
+ \partial_k \hat{W}_{ij} \partial_k V_j
- \partial_k \hat{W} \partial_k V_i\nonumber\\
&& \qquad\quad
+ 2 \partial_j V \partial_i \hat{R}_j
+ 3 V \partial_t V \partial_i V
+ V_j \partial_j V \partial_i V
+ 2 V \partial_j V \partial_i V_j
- \partial_k V_j \partial_i \hat{W}_{jk}\nonumber\\
&& \qquad\quad
+\frac{1}{2} \partial_t V \partial_i \hat{W}
+\frac{1}{2} \partial_j V_j \partial_i \hat{W}
+\frac{1}{2} \hat{W} \partial_{jj} V_i
\biggr\}
+\calO\left(\frac{1}{c^6}\right)\,,\\
\Sigma_{ij} &=& 
\biggl[ 1 +\frac{4V}{c^2}\biggr] \sigma_{ij}
+ \frac{1}{\pi G} \left\{ 
-\frac{1}{8}\delta_{ij} \partial_k V \partial_k V
+\frac{1}{4} \partial_i V \partial_j V
\right\} \nonumber \\ 
&&+\frac{1}{\pi Gc^2}\biggl\{
\left(
-\frac{3}{8}(\partial_t V)^2
- \partial_t V_k \partial_k V
-\frac{1}{2} \partial_k V_l \partial_l V_k
+\frac{1}{2} \partial_k V_l \partial_k V_l
\right)\delta_{ij}
- \partial_k V_i \partial_k V_j \nonumber\\
&& \qquad\quad
+ \partial_t V_j \partial_i V
+ \partial_k V_j \partial_i V_k
+ \partial_t V_i \partial_j V
+ \partial_k V_i \partial_j V_k
- \partial_i V_k \partial_j V_k
\biggr\}
+\calO\left(\frac{1}{c^4}\right)\,,
\end{eqnarray}
\end{subequations}
}\noindent
where we have used the notation $\hat{W}=\hat{W}_{ii}$ and $\hat{Z}=\hat{Z}_{ii}$.
Equivalent expressions for $\Sigma$, $\Sigma_i$ and $\Sigma_{ij}$,
which we have used for testing our calculations, are provided in
Appendix~\ref{appA}.

\section{Application to spinning compact binaries}
\label{III}

\subsection{The pole-dipole effective formalism}
\label{IIIA}

The pole-dipole formalism \cite{Math37, Papa51, Papa51spin,
  CPapa51spin, Tulc1, Tulc2, Traut58, Dixon64, Dixon73, Dixon79, BI80}
is an effective description of point particles endowed with intrinsic
(classical) angular momenta or spins, and moving in an arbitrary
curved background --- in practice the space-time generated by the
particles themselves. The spins can take any orientation and
magnitude, and in particular be close to extremal. In the present work
we shall confine the formalism to terms linear in the spins. At that
level the model can be used for describing black holes as well as
ordinary compact bodies like neutrons stars. Indeed, the internal
structure of the spinning bodies should appear only at the quadratic
level in the spins, \textit{e.g.} through the rotationally induced
quadrupole moment.

The stress-energy tensor of each of the particles is the sum of two
terms, respectively built with a Dirac delta function and a gradient
of a delta function, and integrated over the world line of the
particle, according to:
\begin{equation}\label{Tmunu4D}
T^{\mu\nu}(x) = \sum_{1,2} c^{2} \int^{+\infty}_{-\infty} \ud\tau \;
p^{(\mu}u^{\nu)}\frac{\delta^{(4)}(x-y(\tau))}{\sqrt{-g(x)}} - c
\int^{+\infty}_{-\infty} \ud\tau \; \nabla_{\rho} \left[
  S^{\rho(\mu}u^{\nu)} \frac{\delta^{(4)}(x-y(\tau))}{\sqrt{-g(x)}}
  \right] \;.
\end{equation}
Here the sum is over the two particles, $\tau$ is the proper time
measured along the world line of each particle given by the particle
position $y^\mu(\tau)$; $\delta^{(4)}$ denotes the four-dimensional
Dirac delta function; $u^{\mu}=\ud y^{\mu}/(c\ud\tau)$ is the four
velocity of the particle normalized to $u_\mu u^\mu=-1$; $p^{\mu}$ is
its four linear momentum; and $S^{\mu\nu}$ denotes the antisymmetric
tensor that represents the spin of the particle.\footnote{In our
  convention the spin tensor and all spin variables have the dimension
  of an angular momentum times the speed of light $c$. This is useful
  for counting the effects of spins in the post-Newtonian
  approximation, as the spins appear to be formally ``Newtonian'' for
  maximally spinning particles. All the powers of $1/c$ are kept
  explicitly in our calculations, so that the next-to-next-to-leading
  spin-orbit terms we are looking for here will all carry in front a
  factor $1/c^7$.} Using a $3+1$ space-time split, the particle's
position and coordinate velocity are denoted $y^\mu=(c\,t,
\mathbf{y}(t))$ and $v^\mu(t)=(c, \mathbf{v}(t))$ (where
$v^{\mu}=cu^{\mu}/u^{0}$, with
$u^{0}=[-g_{\rho\sigma}v^{\rho}v^{\sigma}/c^{2}]^{-1/2}$), and we have
\begin{equation}\label{Tmunu3D}
T^{\mu\nu}(\mathbf{x},t) = \sum_{1,2} p^{(\mu}v^{\nu)}
\frac{\delta^{(3)}(\mathbf{x}-\mathbf{y}(t))}{\sqrt{-g(\mathbf{x},t)}}
-\frac{1}{c} \nabla_{\rho} \left[ S^{\rho(\mu}v^{\nu)}
  \frac{\delta^{(3)}(\mathbf{x}-\mathbf{y}(t))}{\sqrt{-g(\mathbf{x},t)}}
  \right]\,,
\end{equation}
where $\delta^{(3)}$ is the three-dimensional Dirac delta function and
the spin tensor $S^{\mu\nu}(t)$ is considered a function of coordinate
time.

Since the spin tensor $S^{\mu\nu}$ has six independent components, one
must impose a supplementary spin condition (SSC) in order to correctly
describe the three independent components of the spin vector. Here we
adopt the covariant SSC (Tulczyjew's condition \cite{Tulc1,Tulc2})
\begin{equation}\label{SSC}
S^{\mu\nu}p_{\nu} = 0\,.
\end{equation}
It can be shown that, with the latter choice for the SSC, the mass
defined by $m^{2}c^{2}=-p^{\mu}p_{\mu}$ as well as the (four-dimensional)
magnitude of the spin defined by $S^{2}=S^{\mu\nu}S_{\mu\nu}/2$, are
conserved along the particle's trajectory. Furthermore the link
between the four velocity $u^{\mu}$ and the four linear momentum
$p^{\mu}$ is entirely specified. However, at linear order in the spins
the linear momentum is simply proportional to the four velocity,
\begin{equation}\label{pu}
p^{\mu}=m\,c \,u^{\mu} + \calO(S^{2})\,,
\end{equation}
so that the covariant SSC reduces to $S^{\mu\nu}u_{\nu} =
\calO(S^{3})$. Furthermore it can be shown that the spin precession
equation reduces to the parallel transport equation
\begin{equation}\label{paral}
	\frac{\uD S^{\mu\nu}}{\ud\tau} = \calO(S^{2})\,,
\end{equation}
where $\uD/\ud\tau$ is the proper time covariant derivative, while the
four dimensional acceleration of the particle is given by
\begin{equation} 
	\frac{\uD u^{\mu}}{\ud \tau} = -\frac{1}{2m c}
        R^{\mu}_{\ph{\mu}\nu\rho\sigma}u^{\nu}S^{\rho\sigma}
        +\calO(S^{2})\,.
\end{equation}
This equation of motion is well known as the Mathisson-Papapetrou
equation \cite{Math37, Papa51, Papa51spin, CPapa51spin}.

\subsection{Multipole moments in the center-of-mass frame}
\label{IIIB}

We work in all intermediate calculations with the spatial components
$S_1^{ij}$ and $S_2^{ij}$ of the two spin tensors ($i,j=1,2,3$),
eliminating the $0i$ components thanks to the covariant SSC
\eqref{SSC}, and then express all our results in terms of some spin
vectors $S_1^i$ and $S_2^i$. Like in Paper II we shall adopt a spin
vector whose magnitude is constant in the usual three-dimensional
Euclidean sense, namely
\begin{equation}\label{const}
\delta_{ij} S_1^i S_1^j = s_1^2\,,\qquad\delta_{ij} S_2^i S_2^j =
s_2^2\,,\qquad s_{1,2} = \text{const}\,.
\end{equation}
We refer to Section IIA in Paper II for the precise construction,
starting from the spatial components of the spin tensors, of spin
vector variables $S_1^i$ and $S_2^i$ with constant magnitude,
Eq.~\eqref{const}. Such spin vectors therefore satisfy
ordinary-looking precession equations,
\begin{equation}\label{precequation}
\frac{\ud S_1^i}{\ud t} = \varepsilon^{ijk}\,\Omega_1^j
\,S_1^k\,,\qquad\frac{\ud S_2^i}{\ud t} = \varepsilon^{ijk}\,\Omega_2^j
\,S_2^k\,.
\end{equation}
All the problem of the evolution of the spins $S_1^i$ and $S_2^i$ in a
binary system reduces to that of finding the ordinary precession
vectors $\Omega_1^i$ and $\Omega_2^i$. Those have been obtained for
the two particles up to 3PN order in Section IIB of Paper II (for spin-orbit effects the precession vectors are independent of the
spins). Now that we have defined the two constant magnitude spins
$S_{1}^i$ and $S_{2}^i$ it will be convenient to express the results
in the center-of-mass frame by means of the variables \cite{K95}
\begin{equation}\label{SSigma}
S^i \equiv S_1^i + S_2^i\,,\qquad\Sigma^i \equiv \frac{m}{m_2}S_2^i -
\frac{m}{m_1}S_1^i\,.
\end{equation}
Here we denote the two individual masses by $m_1$ and $m_2$, the total
mass by $m\equiv m_1+m_2$, and later we shall use the symmetric mass
ratio $\nu\equiv m_1m_2/m^2$ (such that $0<\nu\leqslant 1/4$), and the
mass difference $\delta m \equiv m_1-m_2$.

We compute the spin-orbit terms in the various source multipole
moments which will enter the flux up to 3.5PN order in an arbitrary
frame. At this stage, an interesting check described in Appendix~\ref{appB} can be performed. Next, we adopt the frame of the center-of-mass (CM) defined in
Paper II; notably the particle's trajectories $y_1^i$ and $y_2^i$ in
the CM frame are given in terms of the relative position $x^i\equiv
y_1^i-y_2^i$ and relative velocity $v^i\equiv\ud x^i/\ud
t=v_1^i-v_2^i$ by Eqs.~(3.3) in Paper II. Finally the spin parts of
the multipole moments are expressed in terms of the
conserved-magnitude spins and of the variables \eqref{SSigma}. Our
notation is \textit{e.g.} $(v S)\equiv\mathbf{v}\cdot\mathbf{S}$ for
the ordinary Euclidean scalar product, \textit{e.g.}
$(\mathbf{x}\times\mathbf{\Sigma})^i\equiv\varepsilon^{ijk}x^j\Sigma^k$
for the ordinary cross product, and \textit{e.g.}
$(S,x,v)\equiv\mathbf{S}\cdot
(\mathbf{x}\times\mathbf{v})=\varepsilon^{ijk}S^ix^jv^k$ for the mixed
product. For each source multipole moment, we only present the linear-in-spin part, indicated by the underneath label $S$. We obtain,
extending Ref.~\cite{BBF06} to next-to-next-to-leading order and
writing all terms using our conserved spin variables,
{\allowdisplaybreaks
\begin{subequations}\label{sourcemoments}
\allowdisplaybreaks{
\begin{eqnarray}
\mathop{I}_{S}{}_{ij}&=&
\frac{r\nu}{c^3}\bigg\{
-\frac{8}{3}(\mathbf{S}\times\mathbf{v})^{<i} n^{j>} 
- \frac{8}{3}\frac{\delta m}{m}(\mathbf{\Sigma}\times\mathbf{v})^{<i} n^{j>} \nonumber \\ 
&&\qquad\quad - \frac{4}{3}(\mathbf{n}\times\mathbf{S})^{<i} v^{j>}
 - \frac{4}{3}\frac{\delta m}{m}(\mathbf{n}\times\mathbf{\Sigma})^{<i} v^{j>}\bigg\} \nonumber \\ 
&+& \frac{r\nu}{c^5}\Bigg[\bigg\{
(\mathbf{S}\times\mathbf{v})^{<i} n^{j>}\left(-\frac{26}{21} + \frac{26}{7} \nu\right) v^{2}
+ (\mathbf{\Sigma}\times\mathbf{v})^{<i} n^{j>}\frac{\delta m}{m}\left(-\frac{26}{21} + \frac{116}{21} \nu\right) v^{2} \nonumber \\ 
&& \qquad\quad+ (\mathbf{n}\times\mathbf{S})^{<i} v^{j>}\left(-\frac{4}{21} + \frac{4}{7} \nu\right) v^{2} 
+ (\mathbf{n}\times\mathbf{\Sigma})^{<i} v^{j>}\frac{\delta m}{m}\left(-\frac{4}{21} + \frac{12}{7} \nu\right) v^{2} \nonumber \\ 
&& \qquad\quad+ (\mathbf{S}\times\mathbf{v})^{<i} v^{j>}\left(\frac{4}{21} -\frac{4}{7} \nu\right) (nv) 
+ (\mathbf{\Sigma}\times\mathbf{v})^{<i} v^{j>}\frac{\delta m}{m}\left(\frac{4}{21} -\frac{20}{21} \nu\right) (nv) \nonumber \\ 
&& \qquad\quad+ (n,S,v) v^{<i} v^{j>}\left(-\frac{3}{7} + \frac{9}{7} \nu\right)
+ (n,\Sigma ,v) v^{<i} v^{j>}\frac{\delta m}{m}\left(-\frac{3}{7} + \frac{40}{21} \nu\right)
\bigg\}\nonumber \\
 &&\quad+\frac{Gm}{r}\bigg\{
 (n,S,v) n^{<i} n^{j>}\left(-\frac{38}{21} -\frac{4}{7} \nu\right)
 + (n,\Sigma ,v) n^{<i} n^{j>}\frac{\delta m}{m}\left(-\frac{16}{7} + \frac{26}{21} \nu\right) \nonumber \\ 
&& \qquad\quad+ (\mathbf{n}\times\mathbf{S})^{<i} n^{j>}\left(\frac{17}{21} + \frac{61}{21} \nu\right) (nv) 
+ (\mathbf{n}\times\mathbf{\Sigma})^{<i} n^{j>}\frac{\delta m}{m}\left(1 + \frac{34}{21} \nu\right) (nv) \nonumber \\ 
&& \qquad\quad+ (nS) (\mathbf{n}\times\mathbf{v})^{<i} n^{j>}\left(-2 + \frac{10}{3} \nu\right)
+ (n\Sigma ) (\mathbf{n}\times\mathbf{v})^{<i} n^{j>}\frac{\delta m}{m}\left(-2 + \frac{4}{3} \nu\right) \nonumber \\ 
&& \qquad\quad+ (\mathbf{S}\times\mathbf{v})^{<i} n^{j>}\left(-\frac{11}{7} -\frac{125}{21} \nu\right) 
+ (\mathbf{\Sigma}\times\mathbf{v})^{<i} n^{j>}\frac{\delta m}{m}\left(-\frac{1}{3} -\frac{16}{3} \nu\right) \nonumber \\ 
&& \qquad\quad+ (\mathbf{n}\times\mathbf{S})^{<i} v^{j>}\left(-\frac{22}{3} -\frac{10}{3} \nu\right) 
+ (\mathbf{n}\times\mathbf{\Sigma})^{<i} v^{j>}\frac{\delta m}{m}\left(-\frac{8}{3} -\frac{34}{21} \nu\right)
\bigg\}\Bigg]\nonumber\\
&+&\frac{r\nu}{c^7}\Bigg[\bigg\{
(\mathbf{S}\times\mathbf{v})^{<i} n^{j>}\left(-\frac{58}{63} + \frac{404}{63} \nu -\frac{746}{63} \nu^2\right) v^{4}\nonumber \\ 
&&\qquad\quad + (\mathbf{\Sigma}\times\mathbf{v})^{<i} n^{j>}\frac{\delta m}{m}\left(-\frac{58}{63} + \frac{542}{63} \nu -\frac{1262}{63} \nu^2\right) v^{4} \nonumber \\ 
&&\qquad\quad + (\mathbf{n}\times\mathbf{S})^{<i} v^{j>}\left(-\frac{4}{63} + \frac{23}{63} \nu -\frac{32}{63} \nu^2\right) v^{4} \nonumber \\ 
&&\qquad\quad + (\mathbf{n}\times\mathbf{\Sigma})^{<i} v^{j>}\frac{\delta m}{m}\left(-\frac{4}{63} + \frac{103}{63} \nu -\frac{16}{3} \nu^2\right) v^{4} \nonumber \\ 
&&\qquad\quad + (\mathbf{S}\times\mathbf{v})^{<i} v^{j>}\left(\frac{25}{126} -\frac{173}{126} \nu + \frac{317}{126} \nu^2\right) (nv) v^{2} \nonumber \\ 
&&\qquad\quad + (\mathbf{\Sigma}\times\mathbf{v})^{<i} v^{j>}\frac{\delta m}{m}\left(\frac{25}{126} -\frac{118}{63} \nu + \frac{575}{126} \nu^2\right) (nv) v^{2} \nonumber \\ 
&&\qquad\quad + (n,S,v) v^{<i} v^{j>}\left(-\frac{25}{63} + \frac{179}{63} \nu -\frac{341}{63} \nu^2\right) v^{2} \nonumber \\ 
&&\qquad\quad + (n,\Sigma ,v) v^{<i} v^{j>}\frac{\delta m}{m}\left(-\frac{25}{63} + \frac{76}{21} \nu -\frac{550}{63} \nu^2\right) v^{2}\bigg\}\nonumber \\
 &&\quad+\frac{Gm}{r}\bigg\{
 (n,S,v) n^{<i} n^{j>}\left[\left(-\frac{148}{63} + \frac{542}{63} \nu + \frac{625}{63} \nu^2\right) (nv)^2+\left(\frac{586}{189} -\frac{2354}{189} \nu -\frac{208}{189} \nu^2\right) v^{2}\right] \nonumber \\ 
&&\qquad + (n,\Sigma ,v) n^{<i} n^{j>}\left[\frac{\delta m}{m}\left(-\frac{167}{126} + \frac{53}{14} \nu + \frac{949}{126} \nu^2\right) (nv)^2+\frac{\delta m}{m}\left(-\frac{19}{378} + \frac{27}{14} \nu -\frac{4231}{378} \nu^2\right) v^{2}\right] \nonumber \\ 
&&\qquad + (\mathbf{n}\times\mathbf{S})^{<i} n^{j>}\left[\left(\frac{499}{252} -\frac{1451}{252} \nu -\frac{3487}{252} \nu^2\right) (nv)^3+\left(\frac{85}{252} + \frac{1369}{252} \nu -\frac{3643}{252} \nu^2\right) (nv) v^{2}\right] \nonumber \\ 
&&\qquad + (\mathbf{n}\times\mathbf{\Sigma})^{<i} n^{j>}\left[\frac{\delta m}{m}\left(\frac{253}{252} -\frac{220}{63} \nu -\frac{187}{28} \nu^2\right) (nv)^3+\frac{\delta m}{m}\left(\frac{89}{84} + \frac{11}{14} \nu -\frac{715}{84} \nu^2\right) (nv) v^{2}\right] \nonumber \\ 
&&\qquad + (nS) (\mathbf{n}\times\mathbf{v})^{<i} n^{j>}\left[\left(-\frac{9}{14} + \frac{33}{14} \nu + \frac{89}{14} \nu^2\right) (nv)^2+\left(\frac{5}{42} + \frac{5}{42} \nu -\frac{415}{42} \nu^2\right) v^{2}\right] \nonumber \\ 
&&\qquad + (n\Sigma ) (\mathbf{n}\times\mathbf{v})^{<i} n^{j>}\left[\frac{\delta m}{m}\left(-\frac{9}{14} + \frac{12}{7} \nu + \frac{39}{14} \nu^2\right) (nv)^2+\frac{\delta m}{m}\left(\frac{5}{42} + \frac{5}{21} \nu -\frac{57}{14} \nu^2\right) v^{2}\right] \nonumber \\ 
&&\qquad + (Sv) (\mathbf{n}\times\mathbf{v})^{<i} n^{j>}\left(-\frac{10}{7} + \frac{110}{21} \nu -\frac{10}{7} \nu^2\right) (nv) \nonumber \\ 
&&\qquad + (\Sigma v) (\mathbf{n}\times\mathbf{v})^{<i} n^{j>}\frac{\delta m}{m}\left(-\frac{10}{7} + \frac{80}{21} \nu -\frac{10}{21} \nu^2\right) (nv) \nonumber \\ 
&&\qquad + (\mathbf{S}\times\mathbf{v})^{<i} n^{j>}\left[\left(\frac{2705}{756} -\frac{991}{108} \nu -\frac{9077}{756} \nu^2\right) (nv)^2+\left(-\frac{755}{108} + \frac{7507}{756} \nu + \frac{13655}{756} \nu^2\right) v^{2}\right] \nonumber \\ 
&&\qquad + (\mathbf{\Sigma}\times\mathbf{v})^{<i} n^{j>}\left[\frac{\delta m}{m}\left(\frac{2117}{756} -\frac{14}{3} \nu -\frac{691}{756} \nu^2\right) (nv)^2+\frac{\delta m}{m}\left(-\frac{4637}{756} + \frac{1993}{126} \nu + \frac{17419}{756} \nu^2\right) v^{2}\right] \nonumber \\ 
&&\qquad + (n,S,v) n^{<i} v^{j>}\left(-\frac{284}{189} + \frac{2203}{189} \nu -\frac{3784}{189} \nu^2\right) (nv) \nonumber \\ 
&&\qquad + (n,\Sigma ,v) n^{<i} v^{j>}\frac{\delta m}{m}\left(-\frac{125}{189} + \frac{34}{9} \nu -\frac{410}{189} \nu^2\right) (nv) \nonumber \\ 
&&\qquad + (\mathbf{n}\times\mathbf{S})^{<i} v^{j>}\left[\left(-\frac{95}{126} -\frac{407}{126} \nu + \frac{4175}{126} \nu^2\right) (nv)^2+\left(-\frac{25}{7} + \frac{14}{3} \nu + \frac{128}{21} \nu^2\right) v^{2}\right] \nonumber \\ 
&&\qquad + (\mathbf{n}\times\mathbf{\Sigma})^{<i} v^{j>}\left[\frac{\delta m}{m}\left(-\frac{13}{126} + \frac{122}{63} \nu + \frac{127}{6} \nu^2\right) (nv)^2+\frac{\delta m}{m}\left(-\frac{121}{63} + \frac{482}{63} \nu + \frac{274}{63} \nu^2\right) v^{2}\right] \nonumber \\ 
&&\qquad + (nS) (\mathbf{n}\times\mathbf{v})^{<i} v^{j>}\left(\frac{3}{14} -\frac{11}{14} \nu + \frac{3}{14} \nu^2\right) (nv) \nonumber \\ 
&&\qquad + (n\Sigma ) (\mathbf{n}\times\mathbf{v})^{<i} v^{j>}\frac{\delta m}{m}\left(\frac{3}{14} -\frac{4}{7} \nu + \frac{1}{14} \nu^2\right) (nv) \nonumber \\ 
&&\qquad + (Sv) (\mathbf{n}\times\mathbf{v})^{<i} v^{j>}\left(\frac{5}{7} -\frac{55}{21} \nu + \frac{5}{7} \nu^2\right) \nonumber \\ 
&&\qquad + (\Sigma v) (\mathbf{n}\times\mathbf{v})^{<i} v^{j>}\frac{\delta m}{m}\left(\frac{5}{7} -\frac{40}{21} \nu + \frac{5}{21} \nu^2\right) \nonumber \\ 
&&\qquad + (\mathbf{S}\times\mathbf{v})^{<i} v^{j>}\left(-\frac{191}{378} -\frac{1511}{378} \nu + \frac{1133}{54} \nu^2\right) (nv) \nonumber \\ 
&&\qquad + (\mathbf{\Sigma}\times\mathbf{v})^{<i} v^{j>}\frac{\delta m}{m}\left(-\frac{95}{378} -\frac{262}{63} \nu + \frac{415}{54} \nu^2\right) (nv) \nonumber \\ 
&&\qquad + (n,S,v) v^{<i} v^{j>}\left(-\frac{326}{189} + \frac{55}{189} \nu + \frac{389}{189} \nu^2\right) \nonumber \\ 
&&\qquad + (n,\Sigma ,v) v^{<i} v^{j>}\frac{\delta m}{m}\left(-\frac{506}{189} + \frac{59}{9} \nu + \frac{319}{189} \nu^2\right)\bigg\}\nonumber \\
 &&\quad+\frac{G^{2}m^{2}}{r^2}\bigg\{
(n,S,v) n^{<i} n^{j>}\left(-\frac{2543}{252} -\frac{18083}{252} \nu + \frac{1319}{252} \nu^2\right) \nonumber \\ 
&&\qquad + (n,\Sigma ,v) n^{<i} n^{j>}\frac{\delta m}{m}\left(-\frac{563}{108} -\frac{2951}{84} \nu + \frac{805}{108} \nu^2\right) \nonumber \\ 
&&\qquad + (\mathbf{n}\times\mathbf{S})^{<i} n^{j>}\left(\frac{937}{84} + \frac{12497}{252} \nu + \frac{2039}{252} \nu^2\right) (nv) \nonumber \\ 
&&\qquad + (\mathbf{n}\times\mathbf{\Sigma})^{<i} n^{j>}\frac{\delta m}{m}\left(\frac{3289}{756} + \frac{21145}{756} \nu + \frac{4289}{756} \nu^2\right) (nv) \nonumber \\ 
&&\qquad + (nS) (\mathbf{n}\times\mathbf{v})^{<i} n^{j>}\left(-\frac{1}{3} + \frac{4}{3} \nu + \frac{13}{3} \nu^2\right) \nonumber \\ 
&&\qquad + (n\Sigma ) (\mathbf{n}\times\mathbf{v})^{<i} n^{j>}\frac{\delta m}{m}\left(-\frac{1}{3} -\frac{5}{3} \nu + \frac{4}{3} \nu^2\right) \nonumber \\ 
&&\qquad + (\mathbf{S}\times\mathbf{v})^{<i} n^{j>}\left(\frac{559}{252} + \frac{17693}{252} \nu -\frac{647}{84} \nu^2\right) \nonumber \\ 
&&\qquad + (\mathbf{\Sigma}\times\mathbf{v})^{<i} n^{j>}\frac{\delta m}{m}\left(\frac{4681}{756} + \frac{38987}{756} \nu -\frac{1781}{252} \nu^2\right) \nonumber \\ 
&&\qquad + (\mathbf{n}\times\mathbf{S})^{<i} v^{j>}\left(-\frac{311}{63} + \frac{56}{3} \nu + \frac{13}{63} \nu^2\right) \nonumber \\ 
&&\qquad + (\mathbf{n}\times\mathbf{\Sigma})^{<i} v^{j>}\frac{\delta m}{m}\left(\frac{100}{189} + \frac{3211}{189} \nu + \frac{482}{189} \nu^2\right)\bigg\}\Bigg]\nonumber\\
&+&\mathcal{O}\left(\frac{1}{c^9}\right)\,,
\end{eqnarray}}
\allowdisplaybreaks{
\begin{eqnarray}
\mathop{J}_{S}{}_{ij}&=&
\frac{r\nu}{c}\bigg\{-\frac{3}{2}\Sigma^{<i} n^{j>}\bigg\}\nonumber\\
&+&\frac{r\nu}{c^3}\Bigg[\bigg\{
-\frac{2}{7}\frac{\delta m}{m} v^{2}S^{<i} n^{j>}
+ \Sigma^{<i} n^{j>}\left(-\frac{29}{28} + \frac{143}{28} \nu\right) v^{2} \nonumber \\ 
&&\qquad\quad + \frac{33}{28}\frac{\delta m}{m}(Sv) n^{<i} v^{j>}
+ (\Sigma v) n^{<i} v^{j>}\left(\frac{33}{28} -\frac{155}{28} \nu\right) \nonumber \\ 
&& \qquad\quad+ \frac{3}{7}\frac{\delta m}{m} (nv)S^{<i} v^{j>} 
+ \Sigma^{<i} v^{j>}\left(\frac{3}{7} -\frac{16}{7} \nu\right) (nv) \nonumber \\ 
&& \qquad\quad- \frac{11}{14}\frac{\delta m}{m}(nS) v^{<i} v^{j>}
+ (n\Sigma ) v^{<i} v^{j>}\left(-\frac{11}{14} + \frac{47}{14} \nu\right)\bigg\}\nonumber \\
&&\qquad+\frac{Gm}{r}\bigg\{
-\frac{29}{14}\frac{\delta m}{m}(nS) n^{<i} n^{j>}
+ (n\Sigma ) n^{<i} n^{j>}\left(-\frac{4}{7} + \frac{31}{14} \nu\right) \nonumber \\ 
&&\qquad\quad + \frac{10}{7}\frac{\delta m}{m}S^{<i} n^{j>} + \Sigma^{<i} n^{j>}\left(\frac{61}{28} -\frac{71}{28} \nu\right)\bigg\}\Bigg]\nonumber\\
&+&\frac{r\nu}{c^5}\Bigg[\bigg\{
S^{<i} n^{j>}\frac{\delta m}{m}\left(-\frac{4}{21} + \frac{17}{21} \nu\right) v^{4}
+ \Sigma^{<i} n^{j>}\left(-\frac{253}{336} + \frac{2435}{336} \nu -\frac{5633}{336} \nu^2\right) v^{4} \nonumber \\ 
&&\qquad\quad + (Sv) n^{<i} v^{j>}\frac{\delta m}{m}\left(\frac{269}{336} -\frac{283}{84} \nu\right) v^{2} 
+ (\Sigma v) n^{<i} v^{j>}\left(\frac{269}{336} -\frac{2587}{336} \nu + \frac{6001}{336} \nu^2\right) v^{2} \nonumber \\ 
&&\qquad\quad + S^{<i} v^{j>}\frac{\delta m}{m}\left(\frac{13}{42} -\frac{53}{42} \nu\right) (nv) v^{2} 
+ \Sigma^{<i} v^{j>}\left(\frac{13}{42} -\frac{125}{42} \nu + \frac{293}{42} \nu^2\right) (nv) v^{2} \nonumber \\ 
&&\qquad\quad + (nS) v^{<i} v^{j>}\frac{\delta m}{m}\left(-\frac{41}{84} + \frac{181}{84} \nu\right) v^{2} 
+ (n\Sigma ) v^{<i} v^{j>}\left(-\frac{41}{84} + \frac{55}{12} \nu -\frac{443}{42} \nu^2\right) v^{2} \nonumber \\ 
&&\qquad\quad + (Sv) v^{<i} v^{j>}\frac{\delta m}{m}\left(\frac{5}{84} -\frac{5}{42} \nu\right) (nv) 
+ (\Sigma v) v^{<i} v^{j>}\left(\frac{5}{84} -\frac{10}{21} \nu + \frac{5}{6} \nu^2\right) (nv)
\bigg\}\nonumber \\
 &&\qquad+\frac{Gm}{r}\bigg\{
 (nS) n^{<i} n^{j>}\left[\frac{\delta m}{m}\left(\frac{23}{168} + \frac{415}{84} \nu\right) (nv)^2+\frac{\delta m}{m}\left(-\frac{617}{504} + \frac{809}{252} \nu\right) v^{2}\right] \nonumber \\ 
&&\qquad\quad+ (n\Sigma ) n^{<i} n^{j>}\left[\left(-\frac{13}{168} + \frac{569}{168} \nu -\frac{2147}{168} \nu^2\right) (nv)^2+\left(\frac{229}{504} -\frac{1703}{504} \nu + \frac{2795}{504} \nu^2\right) v^{2}\right] \nonumber \\ 
&&\qquad\quad + (Sv) n^{<i} n^{j>}\frac{\delta m}{m}\left(\frac{331}{252} + \frac{2549}{504} \nu\right) (nv) \nonumber \\ 
&&\qquad\quad + (\Sigma v) n^{<i} n^{j>}\left(-\frac{101}{252} + \frac{449}{72} \nu -\frac{2789}{126} \nu^2\right) (nv) \nonumber \\ 
&&\qquad\quad + S^{<i} n^{j>}\left[\frac{\delta m}{m}\left(-\frac{115}{126} -\frac{487}{126} \nu\right) (nv)^2+\frac{\delta m}{m}\left(\frac{125}{63} -\frac{157}{63} \nu\right) v^{2}\right] \nonumber \\ 
&&\qquad\quad + \Sigma^{<i} n^{j>}\left[\left(\frac{163}{144} -\frac{10529}{1008} \nu + \frac{25247}{1008} \nu^2\right) (nv)^2+\left(-\frac{3175}{1008} + \frac{18413}{1008} \nu + \frac{7009}{1008} \nu^2\right) v^{2}\right] \nonumber \\ 
&&\qquad\quad + (nS) n^{<i} v^{j>}\frac{\delta m}{m}\left(-\frac{317}{126} -\frac{497}{72} \nu\right) (nv) \nonumber \\ 
&&\qquad\quad + (n\Sigma ) n^{<i} v^{j>}\left(-\frac{173}{126} + \frac{2389}{504} \nu + \frac{1073}{126} \nu^2\right) (nv) \nonumber \\ 
&&\qquad\quad + (Sv) n^{<i} v^{j>}\frac{\delta m}{m}\left(\frac{5}{504} + \frac{1649}{504} \nu\right) \nonumber \\ 
&&\qquad\quad + (\Sigma v) n^{<i} v^{j>}\left(\frac{2021}{504} -\frac{3347}{252} \nu -\frac{4127}{504} \nu^2\right) \nonumber \\ 
&&\qquad\quad + S^{<i} v^{j>}\frac{\delta m}{m}\left(\frac{7}{9} + \frac{160}{63} \nu\right) (nv) \nonumber \\ 
&&\qquad\quad + \Sigma^{<i} v^{j>}\left(\frac{131}{504} -\frac{4741}{504} \nu -\frac{5837}{504} \nu^2\right) (nv) \nonumber \\ 
&&\qquad\quad + (nS) v^{<i} v^{j>}\frac{\delta m}{m}\left(\frac{103}{126} -\frac{751}{252} \nu\right) \nonumber \\ 
&&\qquad\quad + (n\Sigma ) v^{<i} v^{j>}\left(-\frac{275}{126} + \frac{1193}{252} \nu + \frac{691}{63} \nu^2\right)\bigg\}\nonumber \\
 &&\qquad+\frac{G^{2}m^{2}}{r^{2}}\bigg\{
 (nS) n^{<i} n^{j>}\frac{\delta m}{m}\left(\frac{407}{126} -\frac{37}{126} \nu\right)
 + (n\Sigma ) n^{<i} n^{j>}\left(\frac{155}{126} -\frac{311}{63} \nu -\frac{877}{252} \nu^2\right) \nonumber \\ 
&&\qquad\quad + S^{<i} n^{j>}\frac{\delta m}{m}\left(-\frac{103}{63} -\frac{25}{63} \nu\right)
+ \Sigma^{<i} n^{j>}\left(-\frac{275}{504} + \frac{3895}{504} \nu + \frac{1571}{504} \nu^2\right)
\bigg\}\Bigg]\nonumber\\
&+&\mathcal{O}\left(\frac{1}{c^7}\right)\,,
\end{eqnarray}}
\allowdisplaybreaks{
\begin{eqnarray}
\mathop{I}_{S}{}_{ijk}&=&
\frac{r^{2}\nu}{c^3}\bigg\{
\frac{9}{2}\frac{\delta m}{m}(\mathbf{S}\times\mathbf{v})^{<i} n^{j} n^{k>}
+ (\mathbf{\Sigma}\times\mathbf{v})^{<i} n^{j} n^{k>}\left(\frac{9}{2} -\frac{33}{2} \nu\right) \nonumber \\ 
&& \qquad\quad+ 3\frac{\delta m}{m}(\mathbf{n}\times\mathbf{S})^{<i} n^{j} v^{k>}
+ (\mathbf{n}\times\mathbf{\Sigma})^{<i} n^{j} v^{k>}\left(3 -9 \nu\right)
\bigg\} \nonumber \\
&+& \frac{r^{2}\nu}{c^5}\Bigg[\bigg\{
(\mathbf{S}\times\mathbf{v})^{<i} n^{j} n^{k>}\frac{\delta m}{m}\left(\frac{41}{20} -\frac{217}{20} \nu\right) v^{2}
+ (\mathbf{\Sigma}\times\mathbf{v})^{<i} n^{j} n^{k>}\left(\frac{41}{20} -21 \nu + \frac{203}{4} \nu^2\right) v^{2} \nonumber \\ 
&& \qquad\quad+ (\mathbf{n}\times\mathbf{S})^{<i} n^{j} v^{k>}\frac{\delta m}{m}\left(\frac{1}{5} -\frac{49}{10} \nu\right) v^{2}
+ (\mathbf{n}\times\mathbf{\Sigma})^{<i} n^{j} v^{k>}\left(\frac{1}{5} -\frac{11}{2} \nu + 19 \nu^2\right) v^{2} \nonumber \\ 
&& \qquad\quad+ (\mathbf{S}\times\mathbf{v})^{<i} n^{j} v^{k>}\frac{\delta m}{m}\left(-\frac{1}{2} + \nu\right) (nv)
+ (\mathbf{\Sigma}\times\mathbf{v})^{<i} n^{j} v^{k>}\left(-\frac{1}{2} + \frac{9}{2} \nu -\frac{17}{2} \nu^2\right) (nv) \nonumber \\ 
&& \qquad\quad+ (n,S,v) n^{<i} v^{j} v^{k>}\frac{\delta m}{m}\left(\frac{7}{10} -\frac{7}{5} \nu\right) 
+ (n,\Sigma ,v) n^{<i} v^{j} v^{k>}\left(\frac{7}{10} -\frac{11}{2} \nu + \frac{19}{2} \nu^2\right) \nonumber \\ 
&& \qquad\quad+ (\mathbf{n}\times\mathbf{S})^{<i} v^{j} v^{k>}\frac{\delta m}{m}\left(\frac{4}{5} -\frac{8}{5} \nu\right) (nv)
+ (\mathbf{n}\times\mathbf{\Sigma})^{<i} v^{j} v^{k>}\left(\frac{4}{5} -4 \nu + 4 \nu^2\right) (nv) \nonumber \\ 
&& \qquad\quad+ (\mathbf{S}\times\mathbf{v})^{<i} v^{j} v^{k>}\frac{\delta m}{m}\left(\frac{7}{10} -\frac{7}{5} \nu\right)
+ (\mathbf{\Sigma}\times\mathbf{v})^{<i} v^{j} v^{k>}\left(\frac{7}{10} -\frac{9}{2} \nu + \frac{13}{2} \nu^2\right)\bigg\}\nonumber \\
 &&\qquad+\frac{Gm}{r}\bigg\{
 (n,S,v) n^{<i} n^{j} n^{k>}\frac{\delta m}{m}\left(\frac{139}{60} + \frac{131}{30} \nu\right)
 + (n,\Sigma ,v) n^{<i} n^{j} n^{k>}\left(\frac{53}{20} -\frac{41}{4} \nu + \frac{9}{4} \nu^2\right) \nonumber \\ 
&& \qquad\quad+ (\mathbf{n}\times\mathbf{S})^{<i} n^{j} n^{k>}\frac{\delta m}{m}\left(-\frac{17}{15} -\frac{479}{60} \nu\right) (nv)
+ (\mathbf{n}\times\mathbf{\Sigma})^{<i} n^{j} n^{k>}\left(-\frac{17}{15} -\frac{1}{12} \nu + \frac{52}{3} \nu^2\right) (nv) \nonumber \\ 
&& \qquad\quad+ (nS) (\mathbf{n}\times\mathbf{v})^{<i} n^{j} n^{k>}\frac{\delta m}{m}\left(\frac{9}{4} -\frac{3}{2} \nu\right)
+ (n\Sigma ) (\mathbf{n}\times\mathbf{v})^{<i} n^{j} n^{k>}\left(\frac{9}{4} -\frac{33}{4} \nu + \frac{9}{4} \nu^2\right) \nonumber \\ 
&& \qquad\quad+ (\mathbf{S}\times\mathbf{v})^{<i} n^{j} n^{k>}\frac{\delta m}{m}\left(\frac{7}{6} + \frac{119}{12} \nu\right)
+ (\mathbf{\Sigma}\times\mathbf{v})^{<i} n^{j} n^{k>}\left(\frac{1}{6} + \frac{119}{12} \nu -\frac{253}{6} \nu^2\right) \nonumber \\ 
&& \qquad\quad+ (\mathbf{n}\times\mathbf{S})^{<i} n^{j} v^{k>}\frac{\delta m}{m}\left(\frac{269}{30} + \frac{257}{30} \nu\right)
+ (\mathbf{n}\times\mathbf{\Sigma})^{<i} n^{j} v^{k>}\left(\frac{89}{30} -\frac{19}{3} \nu -\frac{115}{6} \nu^2\right)\bigg\}\Bigg]\nonumber\\
&+&\mathcal{O}\left(\frac{1}{c^7}\right)\,,
\end{eqnarray}}
\allowdisplaybreaks{
\begin{eqnarray}
\mathop{J}_{S}{}_{ijk}&=&
\frac{r^{2}\nu}{c}\bigg\{
2S^{<i} n^{j} n^{k>}
+ 2\frac{\delta m}{m}\Sigma^{<i} n^{j} n^{k>}
\bigg\} \nonumber\\
&+&\frac{r^{2}\nu}{c^3}\Bigg[\bigg\{
S^{<i} n^{j} n^{k>}\left(\frac{5}{3} -5 \nu\right) v^{2}
+ \Sigma^{<i} n^{j} n^{k>}\frac{\delta m}{m}\left(\frac{5}{3} -\frac{25}{3} \nu\right) v^{2} \nonumber \\ 
&& \qquad\quad + (Sv) n^{<i} n^{j} v^{k>}\left(-\frac{5}{3} + 5 \nu\right)+ (\Sigma v) n^{<i} n^{j} v^{k>}\frac{\delta m}{m}\left(-\frac{5}{3} + \frac{19}{3} \nu\right) \nonumber \\ 
&&  \qquad\quad + S^{<i} n^{j} v^{k>}\left(-\frac{4}{3} + 4 \nu\right) (nv) 
+ \Sigma^{<i} n^{j} v^{k>}\frac{\delta m}{m}\left(-\frac{4}{3} + \frac{14}{3} \nu\right) (nv) \nonumber \\ 
&& \qquad\quad + (nS) n^{<i} v^{j} v^{k>}\left(\frac{4}{3} -4 \nu\right) + (n\Sigma ) n^{<i} v^{j} v^{k>}\frac{\delta m}{m}\left(\frac{4}{3} -\frac{14}{3} \nu\right) \nonumber \\ 
&& \qquad\quad + S^{<i} v^{j} v^{k>}\left(\frac{2}{3} -2 \nu\right) 
+ \Sigma^{<i} v^{j} v^{k>}\frac{\delta m}{m}\left(\frac{2}{3} -\frac{4}{3} \nu\right)\bigg\}\nonumber \\
 &&\qquad+\frac{Gm}{r}\bigg\{
(nS) n^{<i} n^{j} n^{k>}\left(\frac{16}{9} -\frac{16}{3} \nu\right)
+ (n\Sigma ) n^{<i} n^{j} n^{k>}\frac{\delta m}{m}\left(\frac{4}{9} -\frac{11}{9} \nu\right) \nonumber \\ 
&& \qquad\quad+ S^{<i} n^{j} n^{k>}\left(-\frac{10}{3} + 6 \nu\right) 
+ \Sigma^{<i} n^{j} n^{k>}\frac{\delta m}{m}\left(-\frac{10}{3} + \frac{11}{3} \nu\right)
\bigg\}
\Bigg]\nonumber\\
&+&\mathcal{O}\left(\frac{1}{c^5}\right)\,,
\end{eqnarray}}
\allowdisplaybreaks{
\begin{align}
\mathop{I}_{S}{}_{ijkl}=
\frac{r^{3}\nu}{c^3}\bigg\{&
(\mathbf{S}\times\mathbf{v})^{<i} n^{j} n^{k} n^{l>}\left(-\frac{32}{5} + \frac{96}{5} \nu\right)
+ (\mathbf{\Sigma}\times\mathbf{v})^{<i} n^{j} n^{k} n^{l>}\frac{\delta m}{m}\left(-\frac{32}{5} + \frac{84}{5} \nu\right) \nonumber \\ 
& + (\mathbf{n}\times\mathbf{S})^{<i} n^{j} n^{k} v^{l>}\left(-\frac{24}{5} + \frac{72}{5} \nu\right) 
+ (\mathbf{n}\times\mathbf{\Sigma})^{<i} n^{j} n^{k} v^{l>}\frac{\delta m}{m}\left(-\frac{24}{5} + \frac{48}{5} \nu\right)\bigg\}\nonumber\\
+\mathcal{O}&\left(\frac{1}{c^5}\right)\,,
\end{align}}
\allowdisplaybreaks{
\begin{align}
\mathop{J}_{S}{}_{ijkl}=&
\frac{r^{3}\nu}{c}\bigg\{
-\frac{5}{2}\frac{\delta m}{m}S^{<i} n^{j} n^{k} n^{l>}
+ \Sigma^{<i} n^{j} n^{k} n^{l>}\left(-\frac{5}{2} + \frac{15}{2} \nu\right)\bigg\}+\mathcal{O}\left(\frac{1}{c^3}\right)\,.
\end{align}}
\end{subequations}
}\noindent

We were able to perform a few technical tests on this calculation of the
source multipole moments. We verified that the so-called ``surface
terms'' can be computed either by a ``bulk'' integral over the entire
three-dimensional space like other non-compact support terms, or by a
surface integral extending on a sphere at spatial infinity. For such
a test we have to use the alternative form of the blocks $\Sigma$,
$\Sigma_i$ and $\Sigma_{ij}$ given in Eqs.~\eqref{blocks2}; we refer
to Section IVD of Ref.~\cite{BI04mult} for a discussion of this type
of terms. In addition we verified that certain quadratic non-compact
support terms can be alternatively evaluated using some particular
analytic kernels (denoted $Y_L$, $S_L$ and $T_L$ in
Ref.~\cite{BIJ02}). Another, more physical, test of the expressions of
the multipole moments we obtain before going to the CM frame, is the
agreement with the so-called ``boosted Kerr black hole limit'' as
investigated in the Appendix \ref{BBH}.

\subsection{Flux and orbital phasing for circular orbits}
\label{IIIC}

For the spin-orbit effects at the post-Newtonian level considered in
the present paper we can neglect all the corrections
$\mathcal{O}(1/c^5)$ in the relations between the canonical and source
multipole moments, see Eqs.~\eqref{MLSL}
and~\eqref{MijIij}. Furthermore the relations between the radiative
and canonical moments, Eqs.~\eqref{tails}, imply a spin-orbit
contribution due to gravitational wave tails and arising at the 3PN
order; we ignore this contribution here since it has already been
computed in Ref.~\cite{BBF11}, and since the next-to-leading tail
contribution would enter the result at 4PN order only. Finally, for
our present purpose, we can replace all the radiative moments $U_L$
and $V_L$ by the corresponding source moments $I_L$ and $J_L$ up to
the 3.5PN spin-orbit level. We can therefore use for the flux
\eqref{flux} at that order the expression
\begin{align}\label{fluxSO}
\mathcal{F} =&
\frac{G}{c^5}\left\{\frac{1}{5}{I}^{(3)}_{ij}{I}^{(3)}_{ij}
+\frac{1}{c^2}\left[ \frac{1}{189}{I}^{(4)}_{ijk}{I}^{(4)}_{ijk}
  +\frac{16}{45}{J}^{(3)}_{ij}{J}^{(3)}_{ij}\right]
\right.\nonumber\\ &\quad\left.
+\frac{1}{c^4}\left[\frac{1}{9072}{I}^{(5)}_{ijkl}{I}^{(5)}_{ijkl}+\frac{1}{84}{J}^{(4)}_{ijk}{J}^{(4)}_{ijk}\right]
+\frac{1}{c^6}\left[\frac{4}{14175}{J}^{(5)}_{ijkl}{J}^{(5)}_{ijkl}\right]
+ \text{(tails)} + \mathcal{O}\left(\frac{1}{c^8}\right)\right\}\,.
\end{align}
The other terms do not contribute to the spin-orbit effect at the
3.5PN order. We insert the explicit results \eqref{sourcemoments} for
the source multipole moments into Eq.~\eqref{fluxSO}, we compute the
time derivatives using systematically the equations of motion derived
in Papers I \& II, and we specialize the result to the case of
quasi-circular orbits, again using the material from Papers I \& II.

It is useful to introduce an orthonormal moving triad
$\{\mathbf{n},\bm{\lambda},\bm{\ell}\}$ defined by
$\mathbf{n}=\mathbf{x}/r$,
$\bm{\ell}=\bm{L}_\mathrm{N}/\vert\bm{L}_\mathrm{N}\vert$ where
$\bm{L}_\mathrm{N}\equiv m \nu \,\mathbf{x}\times\mathbf{v}$ denotes
the Newtonian orbital angular momentum, and
$\bm{\lambda}=\bm{\ell}\times\mathbf{n}$. Then the spin-orbit
contributions in the flux will depend only on the
projections of the spins perpendicular to the orbital plane, namely
$S_\ell\equiv\bm{\ell}\cdot\mathbf{S}$ and
$\Sigma_\ell\equiv\bm{\ell}\cdot\bm{\Sigma}$, where we recall that
$\mathbf{S}$ and $\bm{\Sigma}$ are defined by
Eqs.~\eqref{SSigma}. Furthermore we denote the relevant post-Newtonian
parameter for circular orbits by
\begin{equation}\label{x}
x = \left(\frac{G\,m\,\omega}{c^3}\right)^{2/3}\,,
\end{equation}
where $\omega$ is the orbital frequency, related to the orbital
separation $r$ by Eq.~(4.2) in Paper II. We are then left with the main
result of the present work, namely the spin-orbit contribution to the
flux up to order 3.5PN, as follows:
\begin{align}
\label{fluxres}
\mathop{\mathcal{F}}_{S} &=\frac{32 c^5}{5
  G}\,x^5\,\nu^2\left(\frac{x^{3/2}}{G\,m^2}\right)\left\{ -4S_\ell
-\frac{5}{4}\frac{\delta m}{m}\Sigma_\ell \right.
\nonumber\\&\left.\qquad+ x \left[
  \left(-\frac{9}{2}+\frac{272}{9}\nu\right)S_\ell
  +\left(-\frac{13}{16}+\frac{43}{4}\nu\right)\frac{\delta
    m}{m}\Sigma_\ell\right]\right.\nonumber\\&\left.\qquad+ x^{3/2}
\left[ -16 \pi\,S_\ell -\frac{31\pi}{6}\,\frac{\delta
    m}{m}\Sigma_\ell\right]\right.  \nonumber\\ &\qquad+ x^2
\left[\left(\frac{476645}{6804}+\frac{6172}{189}\nu
  -\frac{2810}{27}\nu^2\right)S_\ell
  +\left(\frac{9535}{336}+\frac{1849}{126}\nu
  -\frac{1501}{36}\nu^2\right)\frac{\delta m}{m}\Sigma_\ell \right]
\nonumber\\ &\left.\qquad+
\mathcal{O}\left(\frac{1}{c^5}\right)\right\}\,.
\end{align}
We refer to Eq. (231) in \cite{Bliving} for the non-spin part of the energy flux up to the 3.5PN order. The tail-induced spin-orbit effect at 3PN order computed in
Ref.~\cite{BBF11} has also been added, but we recall that we neglect
spin-spin interactions. We have checked that this result is in
complete agreement in the test-mass limit where $\nu\to 0$ with the
result of black-hole perturbation theory on a Kerr background obtained
in Ref.~\cite{TSTS96}.

To obtain the evolution of the orbital phase for quasi-circular orbits
we shall apply the energy conservation balance equation relating the
flux $\mathcal{F}$ to the energy $E$ that is associated with the
conservative part of the equations of motion:
\begin{equation}\label{balance}
\frac{\ud E}{\ud t} = - \mathcal{F} \,.
\end{equation}
Note that the balance equation \eqref{balance} is valid \textit{in
  average} over a long radiation-reaction time scale
$\omega/\dot{\omega}\sim x^{-5/2}=\mathcal{O}(c^5)$; thus short
periodic variations at the orbital frequency $\omega$ and at the spin
precession frequencies $\omega_\text{prec}\sim x\,\omega$ have been
averaged out. In order to apply the balance equation \eqref{balance}
we must ensure that the spins, or rather their projections $S_{1\ell}$
and $S_{2\ell}$ (or equivalently $S_{\ell}$ and $\Sigma_{\ell}$), are secularly constant over the radiation-reaction
time scale $\omega/\dot{\omega}$.

This will be the case of the spin variables with conserved magnitude,
as can be shown either explicitly at a given post-Newtonian order
\cite{W05}, or by the following structural argument valid at linear
order in spins, extending the presentation of Ref.~\cite{BBF11}. In
the center-of-mass frame, the only vectors at our disposal, except for
the spins, are $\mathbf{n}$ and $\mathbf{v}$. Recalling that the spin
vectors are pseudovectors regarding parity transformation, we see that
the only way spin-orbit contributions can enter scalars such as the
energy $E$ or the flux $\mathcal{F}$ is through mixed products
$(n,v,S_1)$ and $(n,v,S_2)$, \textit{i.e.} through the components
$S_{1\ell}$ and $S_{2\ell}$. Now, the same argument applies for the
precession vectors $\bm{\Omega}_{1,2}$ introduced in
Eqs.~\eqref{precequation}: they must be pseudovectors, and, at linear
order in spin, they must only depend on $\bm{n}$ and $\bm{v}$, so that
we must have $\bm{\Omega}_{1} \propto \bm{\ell}$ and $\bm{\Omega}_{2}
\propto \bm{\ell}$; this is explicitly seen for instance in Eq.~(4.5)
of Paper II. Now, the time derivative of the components along
$\bm{\ell}$ of the spins are given by $\ud S_{1\ell}/\ud
t=\mathbf{S}_1\cdot[\ud \bm{\ell}/\ud t+\bm{\ell}\times\bm{\Omega}_1]$
and idem for 2. The second term is zero, and since $\ud \bm{\ell}/\ud
t = \calO(S)$, we obtain that $S_{1\ell}$ and $S_{2\ell}$ are
constants at \emph{linear} order in the spins. This argument is valid at any
post-Newtonian order and for general orbits, but is limited to
spin-orbit terms.

The conservative energy $E$ has been obtained in Paper I and was
reduced to circular orbits in Eq.~(4.6) of Paper II. We recall here
its expression:
\begin{align}
\mathop{E}_{S} &=-\frac{m \nu c^2
  x}{2}\left(\frac{x^{3/2}}{G\,m^2}\right) \left\{ \frac{14}{3}S_\ell
+ 2\frac{\delta m}{m}\Sigma_\ell \right.  \nonumber\\&\left.\qquad+ x
\left[ \left(11-\frac{61}{9}\nu\right)S_\ell
  +\left(3-\frac{10}{3}\nu\right)\frac{\delta
    m}{m}\Sigma_\ell\right]\right.  \nonumber\\&\left.\qquad+ x^2
\left[
  \left(\frac{135}{4}-\frac{367}{4}\nu+\frac{29}{12}\nu^2\right)S_\ell
  +\left(\frac{27}{4}-39\nu+\frac{5}{4}\nu^2\right)\frac{\delta
    m}{m}\Sigma_\ell\right] +
\mathcal{O}\left(\frac{1}{c^5}\right)\right\}\,.
\end{align}
See for instance Eq. (4.6) in Paper II for the complete expression of $E$ including non-spin terms. Applying now the balance equation \eqref{balance}, in which we can
assume by the previous argument that the spin projections $S_\ell$ and
$\Sigma_\ell$ are constant, one obtains the secular decrease of the
orbital frequency as
\begin{align}
\left(\frac{\dot{\omega}}{\omega^2}\right)_{S} &=
\frac{96}{5}\nu\,x^{5/2}\,\left(\frac{x^{3/2}}{G\,m^2}\right)\left\{
-\frac{47}{3}S_\ell -\frac{25}{4}\frac{\delta m}{m}\Sigma_\ell
\right. \nonumber\\ & \qquad+x \left[
  \left(-\frac{5861}{144}+\frac{1001}{12}\nu\right)S_\ell
  +\left(-\frac{809}{84}+\frac{281}{8}\nu\right)\frac{\delta
    m}{m}\Sigma_\ell\right] \nonumber\\ &\left.\qquad+ x^{3/2}
\left[ - \frac{188\pi}{3}\,S_\ell -\frac{151\pi}{6}\,\frac{\delta
    m}{m}\Sigma_\ell\right]\right.  \nonumber\\ &\qquad+ x^2 \left[
  \left(-\frac{4323559}{18144}+\frac{436705}{672}\nu
  -\frac{5575}{27}\nu^2\right)S_\ell\right. \nonumber\\ &\qquad\qquad\qquad\left.\left.
  +\left(-\frac{1195759}{18144} +\frac{257023}{1008}\nu
  -\frac{2903}{32}\nu^2\right)\frac{\delta m}{m}\Sigma_\ell \right]
+\mathcal{O}\left(\frac{1}{c^5}\right)\right\}\,.
\end{align}
Finally by a further integration we obtain the secular evolution of
the orbital phase, or more precisely the so-called ``carrier'' phase
defined by $\phi\equiv\int\omega\,\ud t$, as
\begin{align}
\mathop{\phi}_{S}
&=-\frac{x^{-5/2}}{32\nu}\left(\frac{x^{3/2}}{G\,m^2}\right)\left\{
\frac{235}{6}S_\ell +\frac{125}{8}\frac{\delta m}{m}\Sigma_\ell
\right.  \nonumber\\&\left.\qquad+x \ln x \left[
  \left(-\frac{554345}{2016}-\frac{55}{8}\nu\right)S_\ell
  +\left(-\frac{41745}{448}+\frac{15}{8}\nu\right)\frac{\delta
    m}{m}\Sigma_\ell\right]\right.  \nonumber\\ &\left.\qquad+ x^{3/2}
\left[ \frac{940\pi}{3}\,S_\ell +\frac{745\pi}{6}\,\frac{\delta
    m}{m}\Sigma_\ell\right]\right.  \nonumber\\ &\qquad+ x^2 \left[
  \left(-\frac{8980424995}{6096384}+\frac{6586595}{6048}\nu
  -\frac{305}{288}\nu^2\right)S_\ell\right. \nonumber\\ &\qquad\qquad\qquad\left.\left.+\left(-\frac{170978035}{387072}
  +\frac{2876425}{5376}\nu+\frac{4735}{1152}\nu^2\right)\frac{\delta
    m}{m}\Sigma_\ell \right]
+\mathcal{O}\left(\frac{1}{c^5}\right)\right\}\,.
\end{align}
In the case of precessional binaries, for which the spins are not
aligned or anti-aligned with the orbital angular momentum, the total
phase $\Phi$ is the sum of the latter carrier phase and the
precessional correction arising from the precession of the orbital
plane, $\Phi=\phi+\phi_\text{prec}$. The precessional correction
$\phi_\text{prec}$ can be computed numerically \cite{ACST94} or
analytically (see for instance Ref.~\cite{BBF11} for a computation at
the 1PN order).

\begin{table*}[bh]
\caption{Spin-orbit contributions to the number of gravitational-wave
  cycles $\mathcal{N}_\mathrm{GW} =
  (\phi_\mathrm{max}-\phi_\mathrm{min})/\pi$ accumulated from
  $\omega_\mathrm{min} = \pi\times 10\,\mathrm{Hz}$ to
  $\omega_\mathrm{max} = \omega_\mathrm{ISCO}=c^3/(6^{3/2}G m)$ for
  binaries detectable by ground-based detectors LIGO and VIRGO. For
  each compact object we define the magnitude $\chi_a$ and the
  orientation $\kappa_a$ of the spin by $\mathbf{S}_a\equiv G
  \,m_a^2\,\chi_a\,\hat{\mathbf{S}}_a$ and
  $\kappa_a\equiv\hat{\mathbf{S}}_a \cdot \bm{\ell}$. For comparison,
  we give all the non-spin contributions up to 3.5PN order; however we
  neglect all the spin-spin terms. Notice that these figures are only indicative, and that the relative importance of the different terms changes only slightly when choosing another maximal frequency.
\label{table}}
\begin{center}
{\scriptsize
\begin{tabular}{|r|c|c|c|}
\hline
& $1.4 M_{\odot} + 1.4 M_{\odot}$ & $10 M_{\odot} + 1.4 M_{\odot}$ & $10 M_{\odot} + 10 M_{\odot}$  \\
\hline \hline
Newtonian & $15952.6$ & $3558.9$ & $598.8$ \\
1PN & $439.5$ & $212.4$ & $59.1$ \\
1.5PN & $-210.3+65.6 \kappa_1\chi_1+65.6 \kappa_2\chi_2$ & $-180.9+114.0 \kappa_1\chi_1+11.7 \kappa_2\chi_2$ & $-51.2+16.0 \kappa_1\chi_1+16.0 \kappa_2\chi_2$ \\
2PN & $9.9$ & $9.8$ & $4.0$ \\
2.5PN & $-11.7+9.3 \kappa_1\chi_1+9.3 \kappa_2\chi_2$ & $-20.0+33.8 \kappa_1\chi_1+2.9 \kappa_2\chi_2$ & $-7.1+5.7 \kappa_1\chi_1+5.7 \kappa_2\chi_2$ \\
3PN & $2.6-3.2 \kappa_1\chi_1-3.2 \kappa_2\chi_2$ & $2.3 - 13.2\kappa_1\chi_1 - 1.3 \kappa_2\chi_2$ & $2.2-2.6 \kappa_1\chi_1-2.6 \kappa_2\chi_2$ \\
3.5PN & $-0.9+1.9 \kappa_1\chi_1+1.9 \kappa_2\chi_2$ & $-1.8+11.1 \kappa_1\chi_1+0.8 \kappa_2\chi_2$ & $-0.8+1.7 \kappa_1\chi_1+1.7 \kappa_2\chi_2$ \\
\hline
\end{tabular}
}\end{center}
\end{table*}
As a useful diagnosis to assess the importance of the latter spin
effects, we have computed the number of accumulated gravitational-wave
cycles between some minimal and maximal frequencies, corresponding to
the bandwidth of ground-based detectors.\footnote{Note however that
  the number of cycles of the carrier phase $\phi$ does not reflect
  the precession of the orbital plane, which has to be taken into
  account through $\Phi=\phi+\phi_\text{prec}$.} The results are given
in Table~\ref{table}. They show that the 3.5PN spin-orbit terms computed in the present paper can be numerically larger, for spins close to maximal and for suitable orientations, than the non-spin 3PN or 3.5PN contributions. We thus conclude that they are still relevant to be included in the
gravitational wave templates of LIGO/VIRGO/LISA detectors for an accurate
extraction of the binary parameters.

\section*{Acknowledgements}

It is a pleasure to thank Guillaume Faye for discussions. A.B. is grateful for the support of the European Union FEDER funds,
the Spanish Ministry of Economy and Competitiveness project
FPA2010-16495 and the Conselleria d'Economia Hisenda i Innovacio of
the Govern de les Illes Balears. Our computations were done using the
package xAct, which handles symbolic tensor calculus within the
software Mathematica{\footnotesize \textregistered} \cite{xtensor}.


%
%
\appendix

\section{Alternative expressions for $\Sigma$, $\Sigma_{i}$, $\Sigma_{ij}$}
\label{appA}

In this Appendix we provide alternative expressions for the building
blocks $\Sigma$, $\Sigma_{i}$, $\Sigma_{ij}$, defined by
Eq.~\eqref{Sigma}, which are useful for our practical
computations. They are obtained by rewriting some products of
derivatives so as to make Laplacians appear, for instance using
$2\partial_{i}A\partial_{i}B = \Delta(AB)-A\Delta B-B\Delta A$, and
then using \eqref{defpotentials} to
replace the Laplacians of elementary potentials by source terms. In
this process, new compact-supported terms appear, proportional to
$\sigma$, $\sigma_{i}$, $\sigma_{ij}$. We find
\begin{subequations}\label{blocks2}
\begin{eqnarray}
\Sigma &=& \sigma + \frac{4}{c^{4}} \sigma_{ii} V + \frac{1}{c^{6}}
\left(4 \hat{W}_{ij} \sigma_{ij} + 8 \sigma_{ii} V^2 + 16 V_{i} V_{i}
\sigma\right)-\frac{1}{2} \frac{1}{\pi G c^{2}} \Delta\left[V^2\right]
\nonumber \\ && + \frac{1}{\pi G c^{4}} \left(-\frac{2}{3}
\Delta\left[V^3\right] - \frac{1}{2} \Delta\left[V \hat{W}_{ii}\right]
- \hat{W}_{ij} \partial_{j}\partial_{i}V + 2 \partial_{i}V_{j}
\partial_{j}V_{i} - \frac{1}{2} \left(\partial_{t}V\right)^2 - 2 V_{i}
\partial_{t}\partial_{i}V\right) \nonumber \\ && + \frac{1}{\pi G
  c^{6}} \left(-\frac{2}{3} \Delta\left[V^4\right] - \Delta\left[V^2
  \hat{W}_{ii}\right] + \frac{1}{2} \Delta\left[\hat{W}_{ij}
  \hat{W}_{ij}\right] - \frac{1}{4} \Delta\left[\hat{W}_{ii}
  \hat{W}_{jj}\right] - 4 \Delta\left[V \hat{X}\right]
\right. \nonumber \\ && \qquad \qquad \left. - 2 \Delta\left[V
  \hat{Z}_{ii}\right] - 4 \hat{Z}_{ij} \partial_{j}\partial_{i}V + 8
\partial_{i}V_{j} \partial_{j}\hat{R}_{i} - 8 V_{i} \partial_{j}V_{i}
\partial_{j}V + 2 \partial_{t}V_{i} \partial_{t}V_{i}
\right. \nonumber \\ && \qquad \qquad \left. + 4 \partial_{j}V_{i}
\partial_{t}\hat{W}_{ij} - 2 \partial_{i}V_{i}
\partial_{t}\hat{W}_{jj} - 4 \hat{R}_{i} \partial_{t}\partial_{i}V - 2
\left(\partial_{t}V\right)^2 V - 4 V_{i} \partial_{t}\partial_{i}V V
\right. \nonumber \\ && \qquad \qquad \left. + \frac{1}{2}
\partial_{t}^{2}\hat{W}_{ii} V - 6 V_{i} \partial_{i}V \partial_{t}V -
\partial_{t}\hat{W}_{ii} \partial_{t}V + \frac{1}{2} \hat{W}_{ii}
\partial_{t}^{2}V\right) +\calO\left(\frac{1}{c^8}\right)\,, \\
\Sigma_i &=& \sigma_{i} + \frac{1}{c^{2}} \left(-2 V_{i} \sigma + 2
V \sigma_{i}\right) + \frac{1}{c^{4}} \left(-4 \hat{R}_{i} \sigma + 2
\hat{W}_{ij} \sigma_{j} + 2 V^2 \sigma_{i} + 2 V_{i} \sigma_{jj} + 2
V_{j} \sigma_{ij}\right) \nonumber \\ && + \frac{1}{\pi G c^{2}}
\left(-\frac{1}{2} \Delta\left[V V_{i}\right] + \partial_{j}V
\partial_{i}V_{j} + \frac{3}{4} \partial_{i}V \partial_{t}V\right) +
\frac{1}{\pi G c^{4}} \left(- \Delta\left[\hat{R}_{i} V\right] -
\frac{1}{2} \Delta\left[V^2 V_{i}\right] \right. \nonumber \\ &&
\qquad \qquad \left. - \frac{1}{2} \Delta\left[V_{i}
  \hat{W}_{jj}\right] + \frac{1}{2} \Delta\left[V_{j}
  \hat{W}_{ij}\right] - V_{i} \partial_{j}V \partial_{j}V -
\hat{W}_{jk} \partial_{k}\partial_{j}V_{i} + \partial_{j}\hat{W}_{ik}
\partial_{k}V_{j} \right. \nonumber \\ && \qquad \qquad \left. + 2
\partial_{j}V \partial_{i}\hat{R}_{j} + \frac{3}{2} V_{j}
\partial_{j}V \partial_{i}V - \partial_{k}V_{j}
\partial_{i}\hat{W}_{jk} + \frac{3}{2} V \partial_{i}V \partial_{t}V +
\partial_{t}V \partial_{t}V_{i} \right. \nonumber \\ && \qquad \qquad
\left. + \partial_{j}V \partial_{t}\hat{W}_{ij} - 2 V_{j}
\partial_{t}\partial_{j}V_{i} + \frac{1}{2} V_{i} \partial_{t}^{2}V -
\frac{1}{2} V \partial_{t}^{2}V_{i}\right)
+\calO\left(\frac{1}{c^6}\right)\,,\\ 
\Sigma_{ij} &=& \sigma_{ij}
-\frac{1}{2} \delta^{ij} V \sigma + \frac{1}{c^{2}} \left(2
\delta^{ij} V_{k} \sigma_{k} - 4 V_{(i} \sigma_{j)} + 4 V
\sigma_{ij}\right) + \frac{1}{\pi G} \left(-\frac{1}{16}
\Delta\left[V^2\right] \delta^{ij} + \frac{1}{4} \partial_{i}V
\partial_{j}V\right) \nonumber \\ && + \frac{1}{\pi G c^{2}}
\left(-\frac{1}{2} \Delta\left[V_{i} V_{j}\right] + \frac{1}{4}
\Delta\left[V_{k} V_{k}\right] \delta^{ij} - \frac{3}{8} \delta^{ij}
\left(\partial_{t}V\right)^2 - \frac{1}{2} \delta^{ij}
\partial_{k}V_{l} \partial_{l}V_{k} + 2\partial_{k}V_{(i}
\partial_{j)}V_{k} \right. \nonumber \\ && \qquad \qquad \left. -
\partial_{i}V_{k} \partial_{j}V_{k} - \delta^{ij} \partial_{k}V
\partial_{t}V_{k} + 2\partial_{(i}V \partial_{t}V_{j)} + \frac{1}{8}
\delta^{ij} V \partial_{t}^{2}V\right)
+\calO\left(\frac{1}{c^4}\right)\,.
\end{eqnarray}
\end{subequations}
Such a rewriting allows an independent (and also faster and easier)
calculation for all the ``Laplacian'' terms appearing in the multipole
moments, for which the integrals regularized by means of the Finite
Part operation take a simple form in terms of an angular average at
infinity (see the discussion in Section IVD of Ref.~\cite{BI04mult}).
 
\section{The boosted black hole limit}\label{BBH}
\label{appB}

In contrast to the case of the equations of motion for which we can
perform several crucial verifications (see Paper I), there are not so
many tests one can do in the case of the gravitational waveform and
flux. In addition to the test-mass limit $\nu\to 0$ of the flux
\eqref{fluxres} which as we have seen perfectly recovers the result
from Kerr black hole perturbations \cite{TSTS96}, we can perform
another physical test directly at the level of the multipole moments,
before the reduction to the center-of-mass frame: the so-called
boosted black hole (BBH) limit \cite{BDI04zeta}.

The BBH is obtained in the limiting case when we suppress one of the
two black holes (say 2) by setting its mass $m_2$ and spin $S_2^{ij}$
to be exactly zero into the general expressions of the multipole
moments valid in an arbitrary frame, before going to the CM
frame. What remains are then the multipole moments of a single Kerr
black hole having mass $m_1$ and spin $S_1^{ij}$, and moving with
constant velocity $v_1^i$. In the BBH limit the multipole moments
should agree with those of a single Kerr black hole moving with
constant velocity, \textit{i.e.} a black hole space-time on which a
special Lorentz transformation or boost has been applied. The BBH test
is interesting because it verifies (although only partially) the
global Lorentz invariance of the multipole moments and the radiation
field \cite{BDI04zeta}.

We start with the Kerr metric in harmonic coordinates. Since we are
interested in spin-orbit effects we can work at linear order in the
spin of the black hole. The ``gothic'' metric deviation $H^{\mu\nu}$
of the black hole in the rest frame associated with some harmonic
coordinate system $X^\mu=(cT,\mathbf{X})$ (thus satisfying
$\partial_\nu H^{\mu\nu}=0$) reads
\begin{subequations}\label{Hmunu}\begin{eqnarray}
H^{00} &=&
1-\frac{\left(1+\frac{G\,M}{c^2\,R}\right)^3}{1-\frac{G\,M}{c^2\,R}}\,,\\ H^{0i}
&=& - \frac{2G}{c^4 R^2}\frac{\varepsilon^{ijk}\tilde{\mathcal{S}}_j
  N_k}{1-\frac{G\,M}{c^2\,R}} +
\mathcal{O}(\tilde{\mathcal{S}}^2)\,,\\ H^{ij} &=&
-\frac{G^2\,M^2}{c^4\,R^2}\,N^iN^j + \frac{2G^2M}{c^6
  R^3}N^{(i}\varepsilon^{j)kl}\tilde{\mathcal{S}}_k N_l +
\mathcal{O}(\tilde{\mathcal{S}}^2)\,,\label{Hij}
\end{eqnarray}\end{subequations}
where $M$ is the mass of the black hole and $\tilde{\mathcal{S}}_i$ is
the spin vector of the black hole in the rest frame. For convenience
in this Appendix we define the spin vector directly from the spin
tensor as $\tilde{S}_i\equiv\frac{1}{2}\varepsilon_{ijk}S^{jk}$. In
the rest frame of the black hole we denote the spin tensor by
$\mathcal{S}^{ij}$ and the spin vector appearing in Eqs.~\eqref{Hmunu}
is defined by
$\tilde{\mathcal{S}}_i\equiv\frac{1}{2}\varepsilon_{ijk}\mathcal{S}^{jk}$.
We also denote the radial distance and unit direction in the rest
frame by $R\equiv\vert\mathbf{X}\vert$ and $N^i\equiv
X^i/R$. 

Following Ref.~\cite{BDI04zeta} we apply a global boost
$\Lambda^\mu_{\phantom{\mu}\nu}(\mathbf{V})$ with constant velocity
$\mathbf{V}=(V^i)$. In our conventions we pose
\begin{subequations}\label{Lorentz}
\begin{eqnarray}
\Lambda^0_{\phantom{0}0}(\mathbf{V})&=&
\gamma\,,\\ \Lambda^i_{\phantom{i}0}(\mathbf{V}) &=&
\Lambda^0_{\phantom{0}i}(\mathbf{V})
=\gamma\frac{V^i}{c}\,,\\ \Lambda^i_{\phantom{i}j}(\mathbf{V})&=&
\delta^i_j+\frac{\gamma^2}{\gamma+1}\frac{V^iV_j}{c^2}\,,
\end{eqnarray}
\end{subequations}
with $\gamma\equiv\left(1-\frac{V^2}{c^2}\right)^{-1/2}$. In the
global frame defined by $x^\mu=\Lambda^\mu_{\phantom{\mu}\nu}X^\nu$
the Kerr black hole metric will then be given by
\begin{equation}\label{hmunu}
h^{\mu\nu}(x) =
\Lambda^\mu_{~~\rho}\Lambda^\nu_{~~\sigma}\,H^{\rho\sigma}(\Lambda^{-1}x)\,,
\end{equation}
where
$(\Lambda^{-1})^\mu_{\phantom{\mu}\nu}=\Lambda_\nu^{\phantom{\nu}\mu}$
denotes the inverse Lorentz transformation. The radial distance $r$
and unit direction $n^i$ in the global frame $x^\mu=(ct,\mathbf{x})$
are related to the rest-frame counterparts $R$ and $N^i$ by
\begin{subequations}\label{RNi}\begin{eqnarray}
R&=&r\,\biggl[1+c^2(\gamma^2-1)\left(\frac{t}{r}\right)^2-2\gamma^2
  (Vn)\left(\frac{t}{r}\right)+\gamma^2\frac{(Vn)^2}{c^2}\biggr]^{1/2}\,,\\ N^i
&=&\frac{r}{R}\left[n^i-\gamma
  V^i\left(\frac{t}{r}\right)+\frac{\gamma^2}{\gamma+1}
  \frac{V^i}{c^2}(Vn)\right]\,.
\end{eqnarray}\end{subequations}
See \cite{BDI04zeta}; we denote the usual Euclidean scalar product by
$(Vn)\equiv V^in^i$. In addition the spin must also be transformed and
we find that the spin vector $\tilde{\mathcal{S}}_i$ in the rest frame
is related to the spin vector $\tilde{S}_i$ in the global frame by
\begin{equation}
\tilde{\mathcal{S}}_i = \tilde{S}_i -
\frac{\gamma}{\gamma+1}\frac{(\tilde{S} V)}{c^2}\,V_i\,.
\end{equation}

Notice that the spin vector $\tilde{S}_i$ we are using here should
rather be viewed as a \textit{covector}, since it agrees with the
spatial components of the covariant vector $\tilde{S}_\mu$ satisfying
$u^\mu\tilde{S}_\mu=0$, \textit{i.e.} such that
$\tilde{\mathcal{S}}_0=0$ in the rest frame of the black hole. If one
were to use instead the spin vector $S_i$ with conserved magnitude (as
we did in all of this paper), one should have to apply a correction
which is given at the 2PN order and for the BBH case by
\begin{equation}
\tilde{S}_i= S_i
+\frac{(SV)}{c^2}\left(\frac{1}{2}+\frac{3}{8}\frac{V^2}{c^2}\right)V_i\,.
\end{equation}

We now compute all the required multipole moments of the BBH by
inserting the boosted Kerr metric \eqref{Hmunu}--\eqref{hmunu} into
the general definitions of the source multipole moments. However this
calculation is not straightforward starting from the defining
expressions \eqref{ILJL} of the source multipole moments. Instead it
was found in Ref.~\cite{BDI04zeta} that the best is to use some
different expressions of the multipole moments, entirely given by
surface integrals at spatial infinity; they have been derived in
Eqs.~(2.19) of \cite{BDI04zeta}, see also (2.29) for the practical
implementation. In the present paper we have used the same expressions
of the multipole moments and inserted there the BBH solution
\eqref{Hmunu}--\eqref{hmunu}. Expanding the results at the required
post-Newtonian order, and keeping only the spin parts of the multipole
moments, we obtain:
\begin{subequations}
\begin{eqnarray}\label{IijS}
\mathop{I}_\text{S}{}_{\!ij} &=& \frac{t}{c^3}
\,(\tilde{\mathbf{S}}\times\mathbf{V})_{<i} V_{j>} \left[ - \frac{4}{3}
  - \frac{6}{7} \frac{V^2}{c^2} - \frac{83}{126} \frac{V^4}{c^4}
  \right] + \mathcal{O}\left(\frac{1}{c^9}\right)\,,\\
\mathop{J}_\text{S}{}_{\!ij} &=& \frac{t}{c} \,\tilde{S}_{<i} V_{j>}
\left[ \frac{3}{2} + \frac{17}{28} \frac{V^2}{c^2} + \frac{149}{336}
  \frac{V^4}{c^4} \right] \nonumber\\&+& \frac{t}{c^3} \,V_{<i} V_{j>}
(\tilde{S} V) \left[ - \frac{8}{7} - \frac{2}{3} \frac{V^2}{c^2}
  \right] + \mathcal{O}\left(\frac{1}{c^7}\right)\,,\\
\mathop{I}_\text{S}{}_{\!ijk} &=& \frac{t^2}{c^3}
\,(\tilde{\mathbf{S}}\times\mathbf{V})_{<i} V_{j} V_{k>} \left[ -
  \frac{3}{2} - \frac{5}{4} \frac{V^2}{c^2} \right]+
\mathcal{O}\left(\frac{1}{c^7}\right)\,,\\
\mathop{J}_\text{S}{}_{\!ijk} &=& \frac{t^2}{c} \,\tilde{S}_{<i} V_j
V_{k>} \left[ 2 + \frac{V^2}{c^2} \right] - \frac{4}{3}
\frac{t^2}{c^3} \,V_{<i} V_j V_{k>} (\tilde{S} V) +
\mathcal{O}\left(\frac{1}{c^5}\right)\,,\\
\mathop{I}_\text{S}{}_{\!ijkl} &=& - \frac{8}{5} \frac{t^3}{c^3}
\,(\tilde{\mathbf{S}}\times\mathbf{V})_{<i} V_j V_k V_{l>} +
\mathcal{O}\left(\frac{1}{c^5}\right)\,,\\
\mathop{J}_\text{S}{}_{\!ijkl} &=& \frac{5}{2} \frac{t^3}{c}
\,\tilde{S}_{<i} V_j V_k V_{l>} +
\mathcal{O}\left(\frac{1}{c^3}\right)\,.
\end{eqnarray}
\end{subequations}
These results are in perfect agreement with those obtained directly
from our general computation of the multipole moments of black hole
binaries, by setting $m_2=0$ and $\tilde{S}_{2i}=0$, and making the
identifications $m_1\equiv M$, $\tilde{S}_{1i}\equiv \tilde{S}_i$,
$y_1^i\equiv V^i\,t$ and $v_1^i\equiv V^i$.

\bibliography{ListeRef}


\end{document}